\documentclass[aps,prd,eqsecnum]{revtex4}
\usepackage{latexsym}
\usepackage{bm}

\input epsf
\usepackage{graphicx}
\newcommand{\beq}{\begin{equation}}
\newcommand{\beqa}{\begin{eqnarray}}
\newcommand{\eeq}{\end{equation}}
\newcommand{\eeqa}{\end{eqnarray}}

\begin{document}



\vskip15mm

\title{Second Order Quasi-Normal Mode of the Schwarzschild Black Hole}

\author{Hiroyuki Nakano$^1$ and Kunihito Ioka$^2$}

\affiliation{$^1$Center for Computational Relativity and Gravitation, 
School of Mathematical Sciences, \\
Rochester Institute of Technology, 
Rochester, New York 14623, USA
\\
$^2$Department of Physics, Kyoto University, Kyoto 606-8502, Japan
}

\date{\today}


\begin{abstract}
We formulate and calculate 
the second order quasi-normal modes (QNMs) of a Schwarzschild 
black hole (BH). 
Gravitational wave (GW) from a distorted BH, 
so called ringdown, is well understood as QNMs 
in general relativity. 
Since QNMs from binary BH mergers 
will be detected with high signal-to-noise ratio 
by GW detectors, it is also possible to detect 
the second perturbative order of QNMs, 
generated by nonlinear gravitational interaction near the BH. 
In the BH perturbation approach, 
we derive the master Zerilli equation for the metric perturbation 
to second order and explicitly regularize it 
at the horizon and spatial infinity. 
We numerically solve the second order Zerilli equation 
by implementing the modified Leaver's continued fraction method. 
The second order QNM frequencies are found to be twice 
the first order ones, 
and the GW amplitude is up to $\sim 10\%$ 
that of the first order for the binary BH mergers. 
Since the second order QNMs always exist, 
we can use their detections 
(i) to test the nonlinearity of general relativity, 
in particular the no-hair theorem, 
(ii) to remove fake events in the data analysis of QNM GWs 
and (iii) to measure the distance to the BH.
\end{abstract}


\maketitle

\section{Introduction}

Thanks to the recent technological advance, 
we have almost come to the stage that gravitational waves 
are detectable. In the 21 century, 
the observation of gravitational waves will be
absolutely a new window to our universe 
and also provide a direct experimental test of general relativity. 

There are several on-going projects for the gravitational wave detection 
in the world~\cite{TAMA,Abramovici:1992ah,VIRGO,GEO} 
which are ground-based detectors. Next-generation detectors, 
such as the Large-scale Cryogenic Gravitational wave Telescope 
(LCGT)~\cite{LCGT} in Japan, are also in progress. 
In addition, as for space-based interferometric detectors, 
LISA~\cite{LISA} is now on its R \& D stage 
and DECIGO/BBO~\cite{Seto:2001qf,BBO} is proposed as a future project. 
Since they are space-based observation, 
they will be free from the seismic noise and remarkably sensitive 
to the low frequency gravitational waves below 1 Hz. 

One of the most important gravitational wave sources is 
{\it ringdown} of black holes~\cite{kokkotas99}. 
The black hole perturbation, such as in the late stage of a black hole formation, 
can be described by quasi-normal modes (QNMs) with complex frequencies.
Thus the gravitational radiation is expected as a damped sinusoidal waveform. 
It is important to study QNMs because we can determine the mass and angular 
momentum of a spining black hole by observing the QNM frequencies, 
i.e., the normal-mode frequencies and damping rates. 

For ringdown searches using data of gravitational wave detectors, 
the matched filtering technique is useful 
since these waveform is well understood. 
In the paper~\cite{Tsunesada:2005fe}, a data analysis method to search 
for ringdowns have been discussed 
by using an efficient tiling method for ringdown filters~\cite{Nakano}, 
and an application to the TAMA300 data has been reported. 
Accuracies in the waveform parameter estimations have been found 
that [accuracy of black hole mass] $< 0.9 \%$ 
and [accuracy of Kerr parameter] $< 24 \%$ 
for events with the signal-to-noise ratio (SNR) $\ge 10$~\cite{Tsunesada:2004ft}. 

A promising source that excites QNMs is a merger of binary black holes. 
In these events, we may detect the QNMs with high SNR, 
e.g., the SNR $\sim 10^5$ for $\sim 10^8 M_{\odot}$ black hole mergers 
at $\sim 1$Gpc by LISA~\cite{fh98}, 
since a large fraction of energy ($\sim 1$--$5\% \times$ [total mass]) 
is emitted as gravitational waves of the QNMs. 
Recently, numerical simulations have succeeded in calculating 
the entire phase of BH mergers~\cite{pre05,camp06,baker06}, 
and found that the $\ell=2$, $m=\pm 2$ mode actually 
dominates, carrying away 
$\sim 1$--$5\%$ of the initial rest mass of the system~\cite{Berti:2007fi}. 
The merger rate is also estimated to be large enough~\cite{Enoki:2004ew,Ioka:2005pm}.

Black holes deform appreciably in the merger 
so that the higher-order QNMs could be prominent. 
As an order of magnitude estimate, 
when the gravitational wave energy of ringdown is 
$\sim 1\% \times M$, i.e., 
\begin{eqnarray}
E_{GW} \sim \frac{[\psi^{(1)}]^{2}}{M} \sim 1\% \times M \,,
\end{eqnarray}
where $\psi^{(1)}$ denotes the first order gauge 
invariant waveform function, the dimensionless amplitude of the metric
perturbation is 
\begin{eqnarray}
\frac{\psi^{(1)}}{M} \sim 10\% \,. 
\end{eqnarray}
[See the total radiation energy of the first perturbative order 
in Eq.~(\ref{eq:1st-E}) with $\omega \sim M^{-1}$.]
Hence, the second order amplitude goes up to 
$[\psi^{(1)}/M]^2 \sim 1\%$. 
This means that the second order amplitude 
becomes $\sim 10\%$ of the first order amplitude, 
which is detectable for high SNR events. 
Reduced to its simplest terms, 
the SNR of the second order QNMs is $\sim 10$ 
if the SNR is $\sim 100$ for the first order QNMs.

The essence of higher order QNMs 
has already been discussed in ``Mechanics'' by
Landau \& Lifshitz~\cite{landau76} as an anharmonic oscillation. 
In general, an oscillation with small amplitude $x$ is described by 
an equation, 
\beqa
\ddot x+\omega^2 x=0
\eeqa
with a frequency $\omega$. 
This gives a solution, 
\beqa
x=a \cos(\omega t + \phi) \,,
\eeqa 
where $a$ and $\phi$ denote integration constants. 
Including the second perturbative order with respect to the amplitude, 
the equation has a correction, 
\beqa
\ddot x+\omega^2 x=-\alpha x^2 \,, 
\eeqa
where $\alpha$ is a constant, 
and so the solution is $x=a \cos(\omega t +\phi)+x^{(2)}$, where
\begin{eqnarray}
x^{(2)}=-\frac{\alpha a^2}{2\omega^2}+\frac{\alpha a^2}{6 \omega^2}
\cos 2\omega t \propto a^2 \,.
\end{eqnarray}
Here the first term and second term in the right hand side of the above equation 
have a frequency $(\omega-\omega)$ and $(\omega+\omega)$, respectively. 
The important point is that the second order oscillation always 
exists and is completely determined by the first order one. 
We also note that the frequency does not have a correction at this order. 

In a recent paper~\cite{Ioka:2007ak}, 
we have proposed that higher perturbative order of QNMs, 
generated by nonlinear gravitational interaction near the BH, 
are detectable and worth searching in observations 
and simulations of binary BH mergers. 
In this paper, we give full details 
to calculate the second order QNMs of a black hole. 
Since the previous paper just put forward an order-of-magnitude estimate,
we also perform numerical calculations of the second order QNMs. 
Considering a Schwarzschild black hole as a first step, 
we show that the second order QNMs appear at 
the frequency $\omega^{(2)}=2 \omega^{(1)}$ 
and numerically determine their amplitude 
by the modified Leaver's continued fraction method~\cite{leaver85}. 
Since the second order QNMs always exist, their detections 
can be used as a new test of general relativity. 
Although this paper and the previous paper~\cite{Ioka:2007ak} 
have been the first to study second-order QNMs, 
the second-order analysis of general relativity 
is pioneered by Tomita~\cite{Tomita}, 
and the $\ell=2$, $m=0$ case is studied by 
Gleiser {\it et al}.~\cite{gleiser96,Gleiser:1996yc,gleiser00,nicasio00}. 
It is also extended to 
cosmology~\cite{Acquaviva:2002ud,Matarrese:1997ay,Tomita:2005et,Nakamura:2004rm}.

This paper is organized as follows. 
In Sec.~\ref{sec:general}, 
we consider the second order metric perturbation and 
equations to be satisfied, i.e., the perturbed Einstein equation. 
Our strategy to solve this equation will be given in this section. 
In Sec.~\ref{sec:THE}, 
we summarize the tensor harmonics expansion of the first 
and second order perturbation. 
In Sec.~\ref{sec:1st}, 
we review the Regge-Wheeler-Zerilli formalism~\cite{rw57,z70} 
in the black hole perturbation approach. 
Here, the first order QNMs are also discussed for later use. 
In Sec.~\ref{sec:2nd}, we derive 
the second order Zerilli equation with a source term. 
The source term consists of quadratic terms of 
the first order wave-function. 
In Sec.~\ref{sec:reg}, we regularize the second order source term 
so that it is regular at the horizon and spatial infinity. 
In Sec.~\ref{sec:AF}, 
we discuss how to extract physical information 
from the second order Zerilli function
when we obtain this function. 
In practice, the gauge transformation 
from the Regge-Wheeler gauge to an asymptotic flat gauge 
is considered. 
In Sec.~\ref{sec:num}, we numerically solve 
the second order Zerilli equation and calculate the QNM amplitude 
by implementing a modified Leaver's continued fraction method. 
In Sec.~\ref{sec:dis}, we summarize this paper 
and discuss some remaining problems. 
Some discussions on the first QNMs are given in Appendix~\ref{app:qnm}. 
We clarify the complex nature of QNMs 
and the relation 
between $m$ and $-m$ modes in the spherical harmonics expansion 
in this appendix. 
In this paper, we use units in which $c=G=1$
and follow the conventions 
of Misner {\it et al.}~\cite{Misner:1974qy}
with the signature $-\,+\,+\,+$ for the metric.

\section{Second order metric perturbation}\label{sec:general}

In the black hole perturbation approach, 
we consider second order metric perturbations, 
\begin{eqnarray}
\tilde g_{\mu\nu}=g_{\mu\nu}+h_{\mu\nu}^{(1)}+h_{\mu\nu}^{(2)} \,,
\nonumber 
\end{eqnarray}
with an expansion parameter $A$ which we can identify as a first order amplitude. 
Here, superscripts $(i)$ ($i=1,\,2$) denote the perturbative order, 
i.e., $h_{\mu\nu}^{(1)}$ and $h_{\mu\nu}^{(2)}$ are called the 
first and second order metric perturbations, respectively, 
and $g_{\mu\nu}$ is the background metric. 
In this paper, we consider the Schwarzschild metric 
as the background and use the usual Schwarzschild coordinates, 
\begin{eqnarray}
g_{\mu\nu}dx^{\mu} dx^{\nu} &=&
-f(r)\,dt^2+f(r)^{-1}\, dr^2
+r^2\left(d\theta^2+\sin^2\theta d\phi^2\right) \,;
\nonumber \\ 
f(r) &=& 1-\frac{2\,M}{r} \,.
\end{eqnarray}
Thus, the Greek superscript and subscript indices 
denote $\{t,\,r,\,\theta,\,\phi\}$. 
In the perturbation calculation, we raise and lower all tensor indices 
with this background metric. 

The Einstein tensor $G_{\mu\nu}$ up 
to the second order is formally derived as 
\begin{eqnarray}
G_{\mu\nu}[\tilde g_{\mu\nu}] &=& 
G_{\mu\nu}^{(1)}[h^{(1)}]+G_{\mu\nu}^{(1)}[h^{(2)}]+G_{\mu\nu}^{(2)}[h^{(1)},h^{(1)}]
+O((h^{(1)})^3,\,h^{(1)}h^{(2)},\,(h^{(2)})^2) \,, 
\end{eqnarray}
where we have omitted the spacetime indices $\mu$ and $\nu$ 
of the metric perturbations, 
$h_{\mu\nu}^{(1)}$ and $h_{\mu\nu}^{(2)}$, 
and ignored the third perturbative order, i.e., $O(A^3)$ 
in terms of the expansion parameter. 
$G_{\mu\nu}^{(1)}$ is well known as the linearized Einstein equation, 
\begin{eqnarray}
G_{\mu\nu}^{(1)}[H] &=&
-\frac{1}{2}H_{\mu\nu;\alpha}{}^{;\alpha}+F_{(\mu;\nu)}
-R_{\alpha\mu\beta\nu}H^{\alpha\beta}
-\frac{1}{2}H_{;\mu\nu} -\frac{1}{2} g_{\mu\nu}
(F_{\lambda}{}^{;\lambda}-H_{;\lambda}{}^{;\lambda}) \,; \\ 
F_{\mu} &=& H_{\mu\alpha}{}^{;\alpha} \,.
\nonumber 
\end{eqnarray}
Here, $H_{\mu\nu}$ denotes $h_{\mu\nu}^{(1)}$ or $h_{\mu\nu}^{(2)}$, 
and semicolon ";" in the index 
indicates the covariant derivative with respect to the background metric. 
$G_{\mu\nu}^{(2)}$ consists of quadratic terms in the first order perturbation, 
\begin{eqnarray}
G_{\mu\nu}^{(2)}[h^{(1)},h^{(1)}] &=& 
R_{\mu\nu}^{(2)}[h^{(1)},h^{(1)}] - \frac{1}{2}g_{\mu\nu}R^{(2)}[h^{(1)},h^{(1)}] 
\,; \\
R_{\mu\nu}^{(2)}[h^{(1)},h^{(1)}] 
&=& \frac{1}{4}h^{(1)}_{\alpha\beta;\mu}h^{(1)}{}^{\alpha\beta}{}_{;\nu}
+\frac{1}{2}h^{(1)}{}^{\alpha\beta}(h^{(1)}_{\alpha\beta;\mu\nu}
+h^{(1)}_{\mu\nu;\alpha\beta}
-2h^{(1)}_{\alpha(\mu;\nu)\beta}) \nonumber \\ 
&& 
-\frac{1}{2}(h^{(1)}{}^{\alpha\beta}{}_{;\beta}-\frac{1}{2}h^{(1)}_{\beta}{}^{\beta;\alpha})
(2h^{(1)}_{\alpha(\mu;\nu)}-h^{(1)}_{\mu\nu;\alpha})
+\frac{1}{2}h^{(1)}_{\mu\alpha;\beta}h^{(1)}_{\nu}{}^{\alpha;\beta} 
-\frac{1}{2}h^{(1)}_{\mu\alpha;\beta}h^{(1)}_{\nu}{}^{\beta;\alpha} \,.
\nonumber 
\end{eqnarray}

Since we consider the vacuum Einstein equation now, 
we may solve the following equation for the first perturbative order, 
\begin{eqnarray}
G_{\mu\nu}^{(1)}[h^{(1)}] &=& 0 \,.
\label{eq:formal1st}
\end{eqnarray}
For the second perturbative order, once the first order metric perturbation 
$h^{(1)}$ is obtained, we may solve 
the equation with a source term which can be considered as 
an effective energy momentum tensor. 
\begin{eqnarray}
G_{\mu\nu}^{(1)}[h^{(2)}] &=& - G_{\mu\nu}^{(2)}[h^{(1)},h^{(1)}] \,.
\label{eq:formal2nd}
\end{eqnarray}
Thus, expanding the Einstein's vacuum equation, we can obtain 
basic equations order 
by order~\cite{Bruni:1996im,Nakamura:2003wk,Brizuela:2007we}. 

In the following section, 
we consider the equations in Eqs.~(\ref{eq:formal1st}) and (\ref{eq:formal2nd}) 
by using the tensor harmonics expansion 
which is summarized in the next section. Then, 
the Regge-Wheeler-Zerilli formalism~\cite{rw57,z70} 
is used for coefficients of this expansion. 

In this formalism, the first order QNMs of a black hole is derived from 
the first order Einstein equation~(\ref{eq:formal1st}). 
We discuss the first order QNMs and summarize some formulae
in the Regge-Wheeler-Zerilli formalism in Sec.~\ref{sec:1st}. 
(See also Appendix~\ref{app:qnm}.)

For the second order Einstein equation~(\ref{eq:formal2nd}), 
we also use the same Regge-Wheeler-Zerilli formalism in Sec.~\ref{sec:2nd}, 
while there are some differences. 
The second order equation has a source term which is 
written by quadratic terms of the first order wave-function. 
This source term which is shown in Eq.~(\ref{eq:raw2S}), 
does not behave well at the boundaries. 
The reason of the behavior simply comes from the gauge choice 
and this has no physical meaning. 
Hence, we can regularize this behavior in Sec.~\ref{sec:reg}. 
Based on this regularized source in Eq.~(\ref{eq:Sreg}), 
we numerically compute the second order wave-function in Sec.~\ref{sec:num}. 

In order to extract physical information 
from the first and second order wave-functions, 
we must consider the metric perturbations under 
an asymptotic flat gauge condition. 
The first and second order metric perturbations are obtained 
under the Regge-Wheeler gauge at first, 
but this gauge is not an asymptotic flat one. 
It is necessary to formulate the gauge transformation 
for both the first and second perturbative order. 
This is done in Sec.~\ref{sec:AF}.

\section{Tensor harmonics expansion}\label{sec:THE}

Since the background spacetime has the spherical symmetry, 
all perturbative quantities 
can be expanded by the spherical harmonics $Y_{\ell m}(\theta,\phi)$ 
and its angular derivative. 
In this paper, we consider the following tensor harmonics. 
Almost of all is the same as that of Zerilli's paper~\cite{z70}. 
There are some differences in the notation, therefore we summarize them 
in this section. 

For the first and second order metric perturbations, 
we expand $h_{\mu\nu}^{(i)}$ 
($i=1,\,2$) by tensor harmonics,
\begin{eqnarray}
\bm{h}^{(i)} &=& \sum_{\ell m} \left[
f(r)H^{(i)}_{0\,\ell m}(t,r)\bm{a}_{0\,\ell m}
-i\sqrt{2}H^{(i)}_{1\,\ell m}(t,r)\bm{a}_{1\,\ell m}
+\frac{1}{f(r)}H^{(i)}_{2\,\ell m}(t,r)\bm{a}_{\ell m}
\right. \nonumber \\
& &
-\frac{i}{r}\sqrt{2\ell(\ell+1)}h^{(e)(i)}_{0\,\ell m}(t,r)\bm{b}_{0\,\ell m}
+\frac{1}{r}\sqrt{2\ell(\ell+1)}h^{(e)(i)}_{1\,\ell m}(t,r)\bm{b}_{\ell m}
\nonumber \\
& &
+\sqrt{\frac{1}{2}\ell(\ell+1)(\ell-1)(\ell+2)}G^{(i)}_{\ell m}(t,r)\bm{f}_{\ell m}
+\left(\sqrt{2}K^{(i)}_{\ell m}(t,r)
 -\frac{\ell(\ell+1)}{{\sqrt{2}}}G^{(i)}_{\ell m}(t,r)\right)\bm{g}_{\ell m}
\nonumber \\
& &
\left.
-\frac{\sqrt{2\ell(\ell+1)}}{r}h^{(i)}_{0\,\ell m}(t,r)\bm{c}_{0\,\ell m}
+\frac{i\sqrt{2\ell(\ell+1)}}{r}h^{(i)}_{1\,\ell m}(t,r)\bm{c}_{\ell m}
\right.
\nonumber \\
& &
\left.
+\frac{\sqrt{2\ell(\ell+1)(\ell-1)(\ell+2)}}{2r^2}
h^{(i)}_{2\,\ell m}(t,r)\bm{d}_{\ell m}
\right] \,, \label{eq:hharm}
\end{eqnarray}
where we should be careful not to confuse $G^{(i)}_{\ell m}(t,r)$ 
in the above equation with the perturbed Einstein tensor 
$G_{\mu\nu}^{(i)}$ in the previous section. 
Here, $\bm{a}_{0\,\ell m}$, $\bm{a}_{\ell m}\,,\cdots$ are 
constructed by the spherical harmonics and its derivative. 
The ten tensor harmonics are defined as the following. 
\begin{eqnarray}
{\bm a}_{0\,\ell m}&=&
\left(\begin{array}{cccc}
Y_{\ell m} & 0 & 0 & 0 \\
0 & 0 & 0 & 0 \\
0 & 0 & 0 & 0 \\
0 & 0 & 0 & 0
\end{array}\right) \,,
\label{eq:a0}
\\
{\bm a}_{1\,\ell m}&=&(i/\sqrt{2})
\left(\begin{array}{cccc}
0 & Y_{\ell m} & 0 & 0 \\
Sym & 0 & 0 & 0 \\
0 & 0 & 0 & 0 \\
0 & 0 & 0 & 0
\end{array}\right)\,,
\\
{\bm a}_{\ell m}&=&
\left(\begin{array}{cccc}
0 & 0 & 0 & 0 \\
0 & Y_{\ell m} & 0 & 0 \\
0 & 0 & 0 & 0 \\
0 & 0 & 0 & 0
\end{array}\right)\,,
\\
{\bm b}_{0\,\ell m}&=&
ir[2\ell(\ell+1)]^{-1/2}
\left(\begin{array}{cccc}
0 & 0 & (\partial /\partial \theta)Y_{\ell m}
                             & (\partial /\partial \phi)Y_{\ell m} \\
0 & 0 & 0 & 0 \\
Sym & 0 & 0 & 0 \\
Sym & 0 & 0 & 0
\end{array}\right)\,,
\\
{\bm b}_{\ell m}&=&
r[2\ell(\ell+1)]^{-1/2}
\left(\begin{array}{cccc}
0 & 0 & 0 & 0 \\
0 & 0 & (\partial/\partial \theta)Y_{\ell m}
                              &(\partial /\partial \phi)Y_{\ell m} \\
0 & Sym & 0 & 0 \\
0 & Sym & 0 & 0
\end{array}\right)\,,
\\
{\bm c}_{0\,\ell m}&=&r[2\ell(\ell+1)]^{-1/2}
\left(\begin{array}{cccc}
0 & 0 & (1/\sin \theta)(\partial /\partial \phi)Y_{\ell m}
& -\sin \theta(\partial /\partial \theta)Y_{\ell m} \\
0 & 0 & 0 & 0 \\
Sym & 0 & 0 & 0 \\
Sym & 0 & 0 & 0
\end{array}\right)\,,
\\
{\bm c}_{\ell m}&=&ir[2\ell(\ell+1)]^{-1/2}
\left(\begin{array}{cccc}
0 & 0 & 0 & 0 \\
0 & 0 & (1/\sin \theta)(\partial /\partial \phi)Y_{\ell m}
& -\sin \theta(\partial /\partial \theta)Y_{\ell m} \\
0 & Sym & 0 & 0 \\
0 & Sym & 0 & 0
\end{array}\right)\,,
\\
{\bm d}_{\ell m}&=&-ir^2[2\ell(\ell+1)(\ell-1)(\ell+2)]^{-1/2}
\left(\begin{array}{cccc}
0 & 0 & 0 & 0 \\
0 & 0 & 0 & 0 \\
0 & 0 & -(1/\sin \theta)X_{\ell m} & \sin \theta W_{\ell m} \\
0 & 0 & Sym & \sin \theta X_{\ell m}
\end{array}\right)\,,
\\
{\bm g}_{\ell m}&=&(r^2/\sqrt{2})
\left(\begin{array}{cccc}
0 & 0 & 0 & 0 \\
0 & 0 & 0 & 0 \\
0 & 0 & Y_{\ell m} & 0 \\
0 & 0 & 0 & \sin ^2\theta Y_{\ell m}
\end{array}\right)\,,
\\
{\bm f}_{\ell m}&=&r^2[2\ell(\ell+1)(\ell-1)(\ell+2)]^{-1/2}
\left(\begin{array}{cccc}
0 & 0 & 0 & 0 \\
0 & 0 & 0 & 0 \\
0 & 0 & W_{\ell m} & X_{\ell m} \\
0 & 0 & Sym & -\sin ^2\theta W_{\ell m}
\end{array}\right) \,.
\label{eq:flm}
\end{eqnarray}
Here the $Sym$ denotes components derived from 
the symmetry of the tensors, 
and the angular functions 
$X_{\ell m}$ and $W_{\ell m}$ are given by 
\begin{eqnarray}
X_{\ell m}&=&2{\partial \over \partial \phi}
\left({\partial \over \partial \theta}-\cot \theta \right)Y_{\ell m} \,, 
\label{eq:Xlm}
\\
W_{\ell m}&=&\left({\partial ^2\over \partial \theta ^2}
-\cot \theta {\partial \over \partial \theta}
-{1 \over \sin ^2 \theta}
{\partial ^2 \over \partial \phi ^2}\right) Y_{\ell m}
 \,.
\label{eq:Wlm}
\end{eqnarray}
From the above definition, 
the tensor harmonics can be further classified into even (or polar) 
and odd (or axial) parities. 
Even parity modes are defined by the parity $(-1)^\ell$ 
under the transformation $(\theta,\phi)\to(\pi-\theta,\phi+\pi)$, 
while odd parity modes are by the parity $(-1)^{\ell+1}$. 
Thus, the seven coefficents of the metric perturbation, 
$H^{(i)}_{0\,\ell m},\,H^{(i)}_{1\,\ell m},\,H^{(i)}_{2\,\ell m},\,h^{(e)(i)}_{0\,\ell m},\,
h^{(e)(i)}_{1\,\ell m},\,G^{(i)}_{\ell m}$ and $K^{(i)}_{\ell m}$ 
are called as the even parity part, and the three coefficients, 
$h^{(i)}_{0\,\ell m},\,h^{(i)}_{1\,\ell m}$ and $h^{(i)}_{2\,\ell m}$ 
are the odd parity part.

On the other hand, we consider the right hand side of Eq.~(\ref{eq:formal2nd}) 
as the energy momentum tensor, 
\begin{eqnarray}
{\cal T}_{\mu\nu}^{(2)} &=& 
- \frac{1}{8\pi}G_{\mu\nu}^{(2)}[h^{(1)},h^{(1)}] \,.
\label{eq:calT}
\end{eqnarray}
Then we define the following tensor harmonics expansion, 
\begin{eqnarray}
\bm{{\cal T}}^{(2)} &=& \sum_{\ell m}\left[
{\cal A}_{0\,\ell m}(t,r)\bm{a}_{0\,\ell m}
+{\cal A}_{1\,\ell m}(t,r)\bm{a}_{1\,\ell m}
+{\cal A}_{\ell m}(t,r)\bm{a}_{\ell m}
+{\cal B}_{0\,\ell m}(t,r)\bm{b}_{0\,\ell m}
+ {\cal B}_{\ell m}(t,r)\bm{b}_{\ell m}
 \right. \nonumber \\
& & \hspace{1cm} \left.
+{\cal Q}_{0\,\ell m}(t,r)\bm{c}_{0\,\ell m}
+{\cal Q}_{\ell m}(t,r)\bm{c}_{\ell m}
+{\cal D}_{\ell m}(t,r)\bm{d}_{\ell m}
+{\cal G}_{\ell m}(t,r)\bm{g}_{\ell m}
+{\cal F}_{\ell m}(t,r)\bm{f}_{\ell m}
\right]\,.
\label{eq:Tharm}
\end{eqnarray}
Here the seven coefficents, 
${\cal A}_{0\,\ell m},\,{\cal A}_{1\,\ell m},\,{\cal A}_{\,\ell m},\,{\cal B}_{0\,\ell m},\,
{\cal B}_{\ell m},\,{\cal G}_{\ell m}$ and ${\cal F}_{\ell m}$ 
are the even parity part, and the three coefficients, 
${\cal Q}_{0\,\ell m},\,{\cal Q}_{\ell m}$ 
and ${\cal D}_{\ell m}$ are the odd parity part. 

By using the orthogonality of the above tensor harmonics, 
we can derive the coefficient of the tensor harmonics expansion. 
For example, the coefficient of the energy momentum tensor 
is calculated by 
\begin{eqnarray}
{\cal A}_{0\,\ell m}(t,r) &=& \int \bm{{\cal T}}^{(2)} 
\cdot \bm{a}_{0\,\ell m}^{*}\, d\Omega 
\nonumber \\ 
&=& \int \delta^{\mu \alpha} \delta^{\nu \beta} 
{\cal T}^{(2)}_{\mu\nu} \,a_{0\,\ell m \,\alpha\beta}^{*}\, d\Omega \,,
\end{eqnarray}
where $*$ denotes the complex conjugate, 
$d\Omega=\sin \theta d\theta d\phi$ 
and $\delta^{\mu \alpha}$ 
has the component, $diag(1,\,1,\,1/r^2,\,1/(r^2 \sin^2 \theta))$.

\section{First order Zerilli equation}\label{sec:1st}

We use the Regge-Wheeler-Zerilli 
formalism~\cite{rw57,z70} for the first order metric perturbation 
in the Schwarzschild spacetime. 
There are some reviews about this formalism in~\cite{RWZreview}.
Separating angular variables with tensor harmonics of indices $(\ell,m)$ 
as mentioned in the above section, 
the equations are divided into the even and odd parity parts. 
In this formalism, the master equation for the odd or even parity part arises 
as the Regge-Wheeler or Zerilli equation, respectively. 

Here, we consider only the even parity mode 
in the first order calculation. 
The reason is the following. 
In the case of a head-on collision, 
we have already known that the odd parity perturbation 
does not arise due to symmetry. 
When we treat black hole binaries, 
the radial motion always exists 
in order to merge through the potential barrier of the system. 
Actually the numerical simulations show that
the even parity mode dominates. 
Thus, the above assumption for the first perturbative order 
is adequate as a first step, 
though it may not be the best approximation. 
Furthermore, we should note that it is also sufficient 
to discuss the even parity part for the second order calculation 
under the above assumption. 

In order to discuss the metric perturbations, 
it is necessary to fix the gauge. 
For the first order metric perturbation, 
we impose the Regge-Wheeler (RW) gauge conditions, 
the vanishing of some coefficients of the first order metric perturbation: 
\begin{eqnarray*}
h_{0\,\ell m}^{(e)(1){\rm RW}}=h_{1\,\ell m}^{(e)(1){\rm RW}}=G_{\ell m}^{(1){\rm RW}}=0 
\,. 
\end{eqnarray*}
Here, the suffix ${\rm RW}$ stands for the RW gauge. 
It is noted that in the RW gauge the gauge freedom is completely fixed. 
Although there are seven equations for the even parity part,
introducing the following wave-function, 
\begin{eqnarray}
\psi^{(1)}_{\ell m}(t,r)&=&
\frac{r}{\lambda+1} 
\left[
K_{\ell m}^{(1){\rm RW}} (t,r)
+{\frac { r-2\,M  }
{  \lambda\,r+3\,M  }}
\left(H_{2\,\ell m}^{(1){\rm RW}} ( t,r )
-r\,
{\frac {\partial }{\partial r}}K_{\ell m}^{(1){\rm RW}}(t,r)\right)
\right]
\,;
\label{eq:defpsi}
\\ 
\lambda &=& \frac{(\ell-1)(\ell+2)}{2} \,, \nonumber 
\end{eqnarray}
we can reduce the seven equations to a single equation 
for the function $\psi^{(1)}_{\ell m}$. 
This function $\psi^{(1)}_{\ell m}$ obeys the following Zerilli equation. 
\begin{eqnarray}
\left[-\frac{\partial^2}{\partial t^2}
+\frac{\partial^2}{\partial r_*^2}
-V_{Z}(r)\right] \psi_{\ell m}^{(1)}(t,r) &=& 0 \,; 
\nonumber \\ 
\quad
V_{Z}(r) &=& \left(1-\frac{2M}{r}\right)
\frac{2\lambda^2 (\lambda+1) r^3+6 \lambda^2 M r^2+
18 \lambda M^2 r +18 M^3}{r^3 (\lambda r+3M)^2} \,, 
\label{eq:1stZ}
\end{eqnarray}
where 
\begin{eqnarray}
r_* &=& r+2M \ln \left(\frac{r}{2M}-1\right) \,. 
\nonumber 
\end{eqnarray} 
All the first order metric perturbations can be reconstructed from 
the wave-function $\psi^{(1)}_{\ell m}$. 
The detailed method of the first order $\ell=2$ metric reconstruction 
under the RW gauge condition is summarized in the end 
of this section. 

If $\psi^{(1)}_{\ell m}$ is Fourier analyzed, 
\begin{eqnarray}
\psi^{(1)}_{\ell m}(t,r) &=& 
\int e^{-i\omega t} \psi^{(1)}_{\ell m \omega}(r) d\omega \,, 
\label{eq:fourier-psi}
\end{eqnarray}
the Zerilli equation gives a one-dimensional scattering problem 
with a potential, 
\begin{eqnarray}
\left[\frac{\partial^2}{\partial r_{*}^2} + \omega^2 - V_Z(r)\right]
\psi_{l m \omega}^{(1)}(r)=0.
\end{eqnarray}
Then, the QNMs are obtained by imposing the boundary 
conditions with purely ingoing waves, 
\begin{eqnarray*} 
\psi^{(1)}_{\ell m \omega}(r)e^{-i\omega t} &\sim& e^{-i\omega (t+r_*)} \,,
\end{eqnarray*}
at the horizon of a black hole 
and purely outgoing waves 
\begin{eqnarray*} 
\psi^{(1)}_{\ell m \omega}(r)e^{-i\omega t} \sim e^{-i\omega (t-r_*)} \,,
\end{eqnarray*}
at infinity. 
Such boundary conditions are satisfied at discrete QNM frequencies. 
These frequencies are complex with the real part representing 
the actual frequency of the oscillation, i.e., the normal mode frequency, 
and the imaginary part representing the damping. 
There is an infinite number of QNMs for each harmonic index ($\ell, \,m$)
which are labeled by $n$. Thus, the QNM frequency has 
three indices, ($\ell,\,m,\,n$).
We note that the QNM frequencies have some symmetry shown in 
Eq.~(\ref{eq:QNMsym}). 

In the following, we consider only the $\ell=2, m=\pm 2$ modes 
as the first order perturbations. 
This is because these modes dominate for binary black hole mergers. 
Although we can derive all QNMs without fixing the $m$ mode 
in the case of a Schwarzschild black hole, 
we need to specify $m$ mode 
when we consider the second order perturbations. 
(See Appendix \ref{app:qnm}.)

Furthermore, in this paper, 
we mainly concentrate on
the most long-lived QNM 
in the first perturbative order. 
This mode is characterized by the fundamental ($n=0$) QNM frequency. 
Just for comparison, we will calculate the second order QNM 
in the cases that only the $n=1$ or $2$ QNM 
is excited in the first perturbative order. 
Although the QNM frequencies for $\ell=2$ are given 
in~\cite{leaver85}, 
we calculate them with higher-precision in Table \ref{tab:QNM}
that is necessary to calculate the second perturbative order. 

For later use, we show that 
$\psi^{(1)}_{2-2}$ must be the complex conjugate of $\psi^{(1)}_{22}$
in order to assure that the metric perturbation has a real value. 
For example, the first order metric component $h_{tt}^{(1)RW}$ is 
written by 
\begin{eqnarray}
h_{tt}^{(1)RW} &=& f(r) \left( H_{0\,2 2}^{(1)RW}(t,r) \,Y_{22}(\theta,\phi) 
+ H_{0\,2 -2}^{(1)RW}(t,r)\,Y_{2-2}(\theta,\phi) \right)
\nonumber \\ 
&=& f(r) \left( \hat H_{0\,2 2}^{(1)RW}[\psi^{(1)}_{22}(t,r)] \,Y_{22}(\theta,\phi) 
+ \hat H_{0\,2 -2}^{(1)RW}[\psi^{(1)}_{2-2}(t,r)] \,Y_{2-2}(\theta,\phi) \right)
\nonumber \\ 
&=& f(r) \left( \hat H_{0\,2 2}^{(1)RW}[\psi^{(1)}_{22}(t,r)\,Y_{22}(\theta,\phi)
+\psi^{(1)}_{2-2}(t,r) \,Y_{2-2}(\theta,\phi)] \right) \,,
\end{eqnarray}
where we have used Eqs.~(\ref{eq:hharm}) and (\ref{eq:a0}) 
in the first line, 
and $\hat H_{0\,2 2}^{(1)RW} = \hat H_{0\,2 -2}^{(1)RW}$ 
is a real differential operator with respect to $t$ and $r$ 
defined by Eqs.~(\ref{eq:1stREC}). By using 
\begin{eqnarray}
Y_{22}(\theta,\phi) &=& 
\frac{1}{8}\,\sqrt{\frac {30}{\pi}}
\sin^{2} \theta \ e^{2i\phi}
\,, \nonumber \\ 
Y_{2-2}(\theta,\phi) &=& 
\frac{1}{8}\,\sqrt{\frac {30}{\pi}}
\sin^{2} \theta \ e^{-2i\phi}
\,, 
\label{eq:Y22}
\end{eqnarray}
we can show 
\begin{eqnarray} 
\psi^{(1)}_{22}(t,r) &=& \psi^{(1)*}_{2-2}(t,r) 
\nonumber \\ &\sim& 
A\, e^{-i \omega_{22}^{(1)} (t-r_*)} \quad {\rm for} \ r \to \infty \,, 
\label{eq:1stSET0}
\end{eqnarray}
where the second line represents the most dominant QNM 
(see Appendix~\ref{app:qnm})
and $A$ is some complex number.
Berti {\it et al.}~\cite{Berti:2005ys} have discussed that 
the quasi-normal waveform for each $(\ell,\,m)$ mode 
has four degrees of freedom, i.e., two amplitudes and two phases. 
In this paper, we consider only the even parity mode. 
This means that we need to consider two degrees of freedom 
in our calculation, i.e., the complex number $A$. 

From these wave-functions, 
the reconstruction of the first order metric perturbation 
under the RW gauge is given as the following, 
\begin{eqnarray}
H_{0\,2 \pm2}^{(1)RW}(t,r) &=& 
\hat H_{0\,2 \pm2}^{(1)RW}[\psi^{(1)}_{2 \pm2}(t,r)] 
\nonumber \\ &=& 
{\frac { 2\,{r}^{2}-
2\,rM+3\,{M}^{2}  }{r \left( 2\,r+3\,M \right) }}
{\frac {\partial }{\partial r}} \psi^{(1)}_{2 \pm2} \left( t,r \right) 
-3\,{\frac {  4\,{r}^{3}
+4\,{r}^{2}M+6\,r{M}^{2}+3\,{M}^{3}   }{
{r}^{2} \left( 2\,r+3\,M \right) ^{2}}}\psi^{(1)}_{2 \pm2} 
\left( t,r \right)
\nonumber \\ && 
+ \left( r-2\,M \right)  {\frac {\partial^2 }{\partial r^2}}  
\psi^{(1)}_{2 \pm2}  \left( t,r \right) \,,
\nonumber \\ 
H_{1\,2\pm2}^{(1)RW}(t,r) &=& 
\hat H_{1\,2 \pm2}^{(1)RW}[\psi^{(1)}_{2 \pm2}(t,r)] 
\nonumber \\ &=& 
{\frac {  2\,{r}^{2}-6\,rM-3\,{M}^{2}  }{ \left( 2\,r+3\,M \right)  \left( r-2\,
M \right) }}{\frac {\partial }{\partial t}} 
\psi^{(1)}_{2 \pm2}  \left( t,r \right)
+r {\frac {\partial^2 }{\partial t \partial r}}  \psi^{(1)}_{2 \pm2}  \left( t
,r \right)  \,,
\nonumber \\ 
H_{2\,2 \pm2}^{(1)RW}(t,r) &=& 
\hat H_{0\,2 \pm2}^{(1)RW}[\psi^{(1)}_{2 \pm2}(t,r)] 
\nonumber \\ &=& H_{0\,2 \pm2}^{(1)RW}(t,r) \,,
\nonumber \\
K_{2\pm2}^{(1)RW}(t,r) &=& 
\hat K_{2 \pm2}^{(1)RW}[\psi^{(1)}_{2 \pm2}(t,r)] 
\nonumber \\ &=& 
{\frac { r-2\,M  }{r}}{\frac {\partial }{\partial r}} 
\psi^{(1)}_{2 \pm2}  \left( t,r \right) 
+6\,{\frac {  {r}^{2}+rM+{M}^{2}  }{ \left( 2\,r+3\,M \right) {r}^{2}}}
\psi^{(1)}_{2 \pm2} \left( t,r \right) 
\,.
\label{eq:1stREC}
\end{eqnarray}
In the above equations, we can derive the first order metric perturbation 
only from the wave-functions in simple differential forms. 
This is because there is no source term 
in the vacuum Einstein equation. 
On the contrary, when we construct a second order metric perturbation, 
we need both the second order wave-functions and 
the second order source which have quadratic terms 
of the first order perturbation. 

We use the above metric perturbation under the RW gauge 
to derive the source term of the second order Zerilli equation. 
But, when we extract physical information, 
it is necessary to consider a gauge transformation 
from the RW gauge to an asymptotic flat (AF) gauge. 
This will be discussed in Sec.~\ref{sec:AF}.

\section{Second order Zerilli equation}\label{sec:2nd}

For the second order perturbations, we also separate angular variables 
by tensor harmonics and choose the RW gauge condition. 
First of all, we have considered the first order metric perturbation 
only for the even parity part. The second order source term, 
which is derived from $G_{\mu\nu}^{(2)}[h^{(1)},h^{(1)}]$
in Eq.~(\ref{eq:formal2nd}), 
has also the even parity. 
Therefore, we may discuss the second order metric perturbation 
only for the even parity part, i.e., the Zerilli equation. 

Here if the dominant first order perturbations are the $\ell=2$, $m=\pm 2$ 
even parity part, we are sufficient to consider 
the $\ell=4$, $m=\pm 4$ even parity mode for the second order perturbations 
as shown below. 
The second order source term arises from $G_{\mu\nu}^{(2)}[h^{(1)},h^{(1)}]$. 
This contains products, $(\ell=2,m=2) \times (\ell=2,m=2)$, 
$(\ell=2,m=-2) \times (\ell=2,m=-2)$, 
$(\ell=2,m=-2) \times (\ell=2,m=2)$ and
$(\ell=2,m=2) \times (\ell=2,m=-2)$ of the first order metric perturbation. 
Then the source term has $m=0$ and $\pm 4$ modes 
as a result of the product of the spherical harmonics. 
We can find that the $m=0$ source term does not 
oscillate as a function of time, 
because of the symmetry of the QNM frequencies in Eq.~(\ref{eq:QNMsym}), 
and hence this $m=0$ mode can not be observed as a QNM wave. 
Therefore, we consider only $\ell=4,m=\pm 4$ modes 
in the following calculation. 

In \cite{gleiser96,gleiser00,nicasio00}, 
the second order source term of the $\ell=2$, $m=0$ mode 
which arises from the $\ell=2$, $m=0$ first order perturbations, 
has been discussed. 
These discussion is for example, for the treatment 
in the "close limit approximation" 
of the collision of black holes. 
Their situation is restricted to the axisymmetric case, 
i.e., the $m=0$ mode. Therefore, the second order calculation 
is also only for the $m=0$ mode. In our case, 
the harmonics of the second order metric perturbation 
changes in the $m$ mode as well as in the $\ell$ mode as discussed above. 

Now, we introduce a function for the second perturbative order, 
\begin{eqnarray}
\chi^{(2)}_{4 \pm 4}(t,r)=\frac{r-2M}{3 (3 r+M) }
\left[\frac{r^2}{r-2M} 
\frac{\partial K_{4 \pm 4}^{(2)RW}(t,r)}{\partial t} 
- H_{1\,4 \pm 4}^{(2)RW}(t,r)\right] \,,
\label{eq:defchi}
\end{eqnarray}
where the functions $K_{4 \pm 4}^{(2)RW}$ and 
$H_{1\,4 \pm 4}^{(2)RW}$ are coefficients in Eq.~(\ref{eq:hharm}), i.e., 
derived from the expansion of the second order metric perturbation 
by tensor harmonics as in the first order case. 
We note that the first-order counterpart exactly
satisfies $\chi^{(1)}_{\ell m}=\partial
\psi^{(1)}_{\ell m}/\partial t$. 
Hence, the dimensions are $\psi^{(1)}_{\ell m}(t,r) \sim O(M)$, 
$\chi^{(1)}_{\ell m}(t,r) \sim O(M^0)$ 
and $\chi^{(2)}_{\ell m}(t,r) \sim O(M^0)$. 

Using the function of Eq.~(\ref{eq:defchi}), 
the second order equations for the even parity mode is 
reduced to the Zerilli equation with a second order source term, 
\begin{eqnarray}
\left[-\frac{\partial^2}{\partial t^2}
+\frac{\partial^2}{\partial r_*^2}
-V_{Z}(r)\right] \chi^{(2)}_{4 \pm 4}(t,r) &=& S_{4 \pm 4}(t,r) \,,
\label{eq:Zeqchi}
\end{eqnarray}
where the potential $V_{Z}$ is the same function 
defined in Eq.~(\ref{eq:1stZ}) with $\lambda=9$. 
The source term $S_{4 \pm 4}$ 
is derived as 
\begin{eqnarray}
S_{4 \pm 4}(t,r) &=& 
 {\frac {4\,\pi \,\sqrt {10} \left( r-2\,M \right) ^{2}}
{15(3\,r+M)}}\,{\frac {\partial }{\partial t}}{\cal B}_{4 \pm 4} ( t,r ) 
+ {\frac { 8\,\pi \left( r-2\,M \right) ^{2}}{3 (3\,r+M)}}
\,{\frac {\partial }{\partial t}} {\cal A}_{4 \pm 4} ( t,r ) 
\nonumber \\ && 
- {\frac {4\,\sqrt {2}\,i\pi  \left( r-2\,M \right) ^{2} }
{3 (3\,r+M)}}{\frac {\partial }{\partial r}}{\cal A}_{1\,4 \pm 4} ( t,r ) 
- {\frac {8\,\sqrt {2}\,i\pi  \left( r-2\,M \right)  \left( 4\,r-M \right) M}
{3\,r \left( 3\,r+M \right) ^{2}}}{\cal A}_{1\,4 \pm 4} ( t,r ) 
\nonumber \\ && 
- {\frac {4\,\sqrt {10} \,i\pi \left( r-2\,M \right) ^{2}}{15(3\,r+M)}}
{\frac {\partial }{\partial r}}{\cal B}_{0\,4 \pm 4} ( t,r ) 
- \frac{8\,\sqrt {5}\, \pi \left( r-2\,M \right) }{15}
{\frac {\partial }{\partial t}}{\cal F}_{4 \pm 4} ( t,r ) 
\nonumber \\ && 
+ {\frac {4\,\sqrt {10}\,i\pi \,
\left( 4\,{M}^{2}+7\,M\,r+27\,{r}^{2} \right)  \left( r-2\,M \right) }
{15\,r \left( 3\,r+M \right) ^{2}}} {\cal B}_{0\,4 \pm 4} ( t,r ) 
\,,
\label{eq:2ndS4m}
\end{eqnarray}
where the functions ${\cal B}_{4 \pm4}$ etc. 
are coefficients of the tensor harmonics expansion 
of $G_{\mu\nu}^{(2)}[h^{(1)},h^{(1)}]$
in Eqs.~(\ref{eq:calT}) and (\ref{eq:Tharm}). 
Since these coefficients are written by 
quadratic terms of 
the first order metric perturbation, 
we can rewrite the second order source term 
$S_{4 \pm 4}$  in terms of $\psi^{(1)}_{2\pm 2}$ 
by using Eq.~(\ref{eq:1stREC}) as 
\begin{eqnarray}
S_{4 \pm 4}(t,r) &=&
\frac{r-2\,M}{42}\, \frac{\sqrt{70}}{\sqrt {\pi }}
\Biggl\{ 
-\frac{3}{{r}^{5} \left( 3\,r+M \right) ^{2} \left( 2\,r+3\,M
 \right) ^{3}}
\, \left( 24
\,{r}^{7}+1120\,{r}^{6}M+1052\,{r}^{5}{M}^{2}-798\,{r}^{4}{M}^{3}
\right.
\nonumber \\ && \left.
-2586
\,{r}^{3}{M}^{4}-2396\,{r}^{2}{M}^{5}-270\,r{M}^{6}-9\,{M}^{7}
 \right)  \left( {\frac {\partial }{\partial t}}\psi^{(1)}_{2\pm 2} \left( t,r
 \right)  \right) {\frac {\partial }{\partial r}}\psi^{(1)}_{2\pm 2} \left( t,r
 \right)
\nonumber \\ && 
+{\frac { 1}{{r}^{4} \left( 3\,r+M \right) ^{2} \left( 2\,r+3
\,M \right) ^{2}}}
\left( 132\,{r}^{6} -136\,M{r}^{5}+994\,{r}^{4}{M}^{2}-378\,{r}^{3}{M}^{3}
-2306\,{r}^{2}{M}^{4}
\right.
\nonumber \\ && \left. 
-270\,r{M}^{5}-9\,{M}^{6}
\right)  \left( {\frac {\partial ^{2}}{\partial r\partial t}}
\psi^{(1)}_{2\pm 2} \left( t,r \right)  \right) 
{\frac {\partial }{\partial r}}\psi^{(1)}_{2\pm 2}
 \left( t,r \right) 
\nonumber \\ && 
+{\frac { \left( 78\,{r}^{4}-128\,{r}^{3}M-264\,{r}^
{2}{M}^{2}-100\,r{M}^{3}+3\,{M}^{4} \right)  }{r \left( 2\,r+3\,M
 \right)  \left( 3\,r+M \right) ^{2} \left( r-2\,M \right) }}
\left( {\frac {\partial ^
{3}}{\partial {t}^{3}}}\psi^{(1)}_{2\pm 2} \left( t,r \right)  \right) {\frac {
\partial }{\partial r}}\psi^{(1)}_{2\pm 2} \left( t,r \right) 
\nonumber \\ && 
+{\frac {
 \left( 84\,{r}^{5}-122\,{r}^{4}M-281\,{r}^{3}{M}^{2}-6\,{r}^{2}{M}^{3
}+237\,r{M}^{4}+33\,{M}^{5} \right)  }{{r}^{2} \left( 2\,r+3\,M
 \right)  \left( r-2\,M \right) ^{2} \left( 3\,r+M \right) ^{2}}}
\left( {\frac {\partial ^{2}}{
\partial {t}^{2}}}\psi^{(1)}_{2\pm 2} \left( t,r \right)  \right) {\frac {\partial }{
\partial t}}\psi^{(1)}_{2\pm 2} \left( t,r \right) 
\nonumber \\ && 
-{\frac { \left( 66\,{r}^{4}-106\,{r}^{3}M-220\,{r}^{2}{M}^{2}-156\,r{M}
^{3}-45\,{M}^{4} \right)  }{r \left( 2\,r+3\,M
 \right)  \left( 3\,r+M \right) ^{2} \left( r-2\,M \right) }}
\left( {\frac {\partial ^{3}}{\partial t^2
\partial r}}\psi^{(1)}_{2\pm 2} \left( t,r \right)  \right) {\frac {
\partial }{\partial t}}\psi^{(1)}_{2\pm 2} \left( t,r \right) 
\nonumber \\ && 
-{\frac { 9}{{r}^{6}
 \left( 3\,r+M \right) ^{2} \left( 2\,r+3\,M \right) ^{4}}}
\left( 336\,{r}^{8}+3664\,{r}^{7}M+10144\,{r}^{6}{M}^{2}+15052
\,{r}^{5}{M}^{3}
\right.
\nonumber \\ && \left.
+13444\,{r}^{4}{M}^{4}+7386\,{r}^{3}{M}^{5}+2648\,{r}^
{2}{M}^{6}+270\,r{M}^{7}+9\,{M}^{8} \right) \psi^{(1)}_{2\pm 2} \left( t,r \right) {
\frac {\partial }{\partial t}}\psi^{(1)}_{2\pm 2} \left( t,r \right) 
\nonumber \\ && 
+2\,{\frac 
{ \left( r-2\,M \right)  \left( 3\,r+2\,M \right)  }{
r \left( 3\,r+M \right) ^{2}}} \left( {\frac {
\partial ^{2}}{\partial {t}^{2}}}\psi^{(1)}_{2\pm 2} \left( t,r \right)  \right) {
\frac {\partial ^{2}}{\partial r\partial t}}\psi^{(1)}_{2\pm 2} \left( t,r \right)
-3\, \left( {\frac {\partial ^{3}}{
\partial t^2\partial r}}\psi^{(1)}_{2\pm 2} \left( t,r \right)  \right) {
\frac {\partial ^{2}}{\partial r\partial t}}\psi^{(1)}_{2\pm 2} \left( t,r \right) 
\nonumber \\ && 
-{\frac { 3 }{{r}^{5} \left( 3\,r+M
 \right) ^{2} \left( 2\,r+3\,M \right) ^{3}}}
\left( 72\,{r}^{7}+1240\,{r}^{6}M+944\,{r}^{5}{M}^{2}-1284
\,{r}^{4}{M}^{3}
\right.
\nonumber \\ && \left.
-2910\,{r}^{3}{M}^{4}-2396\,{r}^{2}{M}^{5}-270\,r{M}^{6}
-9\,{M}^{7} \right) \psi^{(1)}_{2\pm 2} \left( t,r \right) {\frac {\partial ^{2}}{
\partial r\partial t}}\psi^{(1)}_{2\pm 2} \left( t,r \right)
\nonumber \\ && 
+3\,{\frac {{r}^{2} }{ \left( r-2\,M \right) ^{2}}}
\left( {\frac {\partial ^{3}}{\partial {t}^{3}}}\psi^{(1)}_{2\pm 2} \left( t,r
 \right)  \right) {\frac {\partial ^{2}}{\partial {t}^{2}}}\psi^{(1)}_{2\pm 2}
 \left( t,r \right) 
\nonumber \\ && 
+3\,{\frac {
 \left( 132\,{r}^{5}+196\,{r}^{4}M+174\,{r}^{3}{M}^{2}+114\,{r}^{2}{M}
^{3}+28\,r{M}^{4}-3\,{M}^{5} \right) }{{r}^{2}
 \left( 2\,r+3\,M \right) ^{2} \left( 3\,r+M \right) ^{2} \left( r-2\,
M \right) }}\psi^{(1)}_{2\pm 2} \left( t,r \right) {\frac {
\partial ^{3}}{\partial {t}^{3}}}\psi^{(1)}_{2\pm 2} \left( t,r \right)  
\Biggr\} \,.
\label{eq:raw2S}
\end{eqnarray}

The second order metric perturbations 
can be reconstructed from $\chi^{(2)}_{4 \pm 4}$ 
under the RW gauge as 
\begin{eqnarray}
{\frac {\partial }{\partial t}}K_{4 \pm 4}^{(2)RW}
\left( t,r \right) &=& 
{\frac { 30\,{r}^{2}+9\,rM+2\,{M}^{2}  }
{{r}^{2} \left( 3\,r+M \right) }}\chi^{(2)}_{4 \pm 4}\left( t,r \right)
+{\frac { r-2\,M  }{r}}
{\frac {\partial }{\partial r}}\chi^{(2)}_{4 \pm 4}\left( t,r \right)
\nonumber \\ && 
+\frac{4\,\sqrt {2}}{3}\,
{\frac {i
\pi \,r \left( r-2\,M \right) }{3\,r+M}}{\cal A}_{1\,4 \pm 4} \left( t,r \right) 
+\frac{4\sqrt {5}}{15}\,{\frac {i\pi \,r
 \left( r-2\,M \right) }{3\,r+M}} {\cal B}_{0\,4 \pm 4} \left( t,r \right) 
\,, 
\nonumber \\ 
{\frac {\partial }{\partial t}}H_{2\,4 \pm 4}^{(2)RW}\left( t,r \right) &=& 
r{\frac {
\partial ^{2}}{\partial r\partial t}}K_{4 \pm 4}^{(2)RW} \left( t,r \right) 
+3\,{\frac {M}{{r}^{2}}}\chi^{(2)}_{4 \pm 4}\left( t,r \right) 
-3\,{\frac { 3\,r+M  }{r}}{\frac {\partial }{\partial r}}\chi^{(2)}_{4 \pm 4}\left( t,r \right)
-\frac{4\,\sqrt {5}}{5}\,i\pi \, r\,{\cal B}_{0\,4 \pm 4} \left( t,r \right)
\,, 
\nonumber \\ 
H_{1\,4 \pm 4}^{(2)RW}\left( t,r \right) &=& 
-3\,{\frac { 3\,r+M  }{r-2\,M}}\chi^{(2)}_{4 \pm 4} \left( t,r \right) 
+{\frac {{r}^{2}}{r-2\,M}}{\frac {\partial }{\partial t}}
K_{4 \pm 4}^{(2)RW} \left( t,r \right) 
\,, \nonumber \\ 
H_{0\,4 \pm 4}^{(2)RW}\left( t,r \right) &=& 
H_{2\,4 \pm 4}^{(2)RW}\left( t,r \right) 
+{\frac {8\,\sqrt {5}}{15}}\,\pi \,{r}^{2}\,{\cal F}_{4 \pm 4} \left( t,r \right) 
\,.
\label{eq:rec2nd}
\end{eqnarray}
Unlike the first order metric reconstruction, 
we need the information derived from $G_{\mu\nu}^{(2)}[h^{(1)},h^{(1)}]$, 
such as ${\cal A}_{1\,4 \pm 4}$ and ${\cal B}_{0\,4 \pm 4}$ 
in Eq.~(\ref{eq:Tharm}).

\section{Regularization of source term}\label{sec:reg}

Although 
we show the second order $\ell=4$, $m=\pm 4$ source term $S_{4 \pm 4}$ 
in terms of $\psi^{(1)}_{2\pm 2}$ in Eq.~(\ref{eq:raw2S}), 
the raw expression $S_{4 \pm 4}$ does not behave well at infinity. 
This is not suitable for numerical calculations. 
We can find $S_{4 \pm 4} \sim O(r^0)$ at infinity 
because at large $r$ the first order wave-function behaves as
\begin{eqnarray}
\psi^{(1)}_{2\pm 2}(t,r)=\frac{1}{3}F_I''(t-r_*)+\frac{1}{r}F_I'(t-r_*)
+\frac{1}{r^2}\left[F_I(t-r_*)-MF_I'(t-r_*)\right]+O(r^{-3}) \,,
\label{eq:asympt1st}
\end{eqnarray}
where $F_I$ is some function of $(t-r_*)$ 
and $F_I'(x)$ denotes $d F_I(x)/dx$. 
In order to obtain a finite solution $\chi^{(2)}_{4 \pm 4}$, 
we have to make at least the second order source term $\sim O(r^{-2})$ 
by some regularization, which 
is the same order of the potential $V_Z \sim O(r^{-2})$ 
in Eqs.~(\ref{eq:Zeqchi}) and (\ref{eq:1stZ}). 
On the other hand, the second order source behaves well at the horizon, 
i.e., $\sim O(r-2M)$, with 
\begin{eqnarray}
\psi^{(1)}_{2\pm 2}(t,r)=F_H'(t+r_*)+\frac{1}{4}\frac{F_H(t+r_*)}{M}
+\frac{27}{56}\frac{F_H(t+r_*)}{M^2}(r-2M)+O[(r-2M)^2] \,,
\end{eqnarray}
where $F_H$ is some function of $(t+r_*)$. 

We can regularize the second order source term 
by introducing the following regularized function,
\begin{eqnarray}
\chi^{(2)\,reg}_{4 \pm 4}(t,r) &=& 
\chi^{(2)}_{4 \pm 4}(t,r)
-\zeta^{(2)}_{4 \pm 4}(t,r) \,;
\nonumber \\ 
\zeta^{(2)}_{4 \pm 4}(t,r) &=& 
\frac{\sqrt{70}}{126\sqrt{\pi}}
\frac{(r-2M)^2}{r}
\left(\frac{\partial}{\partial r} \psi^{(1)}_{2\pm 2}(t,r)\right)
\frac{\partial^2}{\partial r \partial t}
\psi^{(1)}_{2\pm 2}(t,r) \,.
\label{eq:chireg}
\end{eqnarray}
The regularized function $\chi^{(2)\,reg}_{4 \pm 4}$ 
satisfies the Zerilli equation (\ref{eq:Zeqchi}) 
with a well-behaved source term, i.e., 
\begin{eqnarray}
\left[-\frac{\partial^2}{\partial t^2}
+\frac{\partial^2}{\partial r_*^2}
-V_{Z}(r)\right] \chi^{(2)\,reg}_{4 \pm 4}(t,r) &=& 
S_{4 \pm 4}(t,r)-\left[-\frac{\partial^2}{\partial t^2}
+\frac{\partial^2}{\partial r_*^2}
-V_{Z}(r)\right] \zeta^{(2)}_{4 \pm 4}(t,r) 
\nonumber \\ 
&=& S_{4 \pm 4}^{reg}(t,r) \,,
\label{eq:Zeqchi-reg}
\end{eqnarray}
where $S_{4 \pm 4}^{reg} \sim O(r^{-2})$ at infinity and 
$S_{4 \pm 4}^{reg} \sim O(r-2M)$ at the horizon. 
Thus we can remove an unphysical gauge-dependent divergence. 

It should be noted that such a regularization is not unique, 
and for example, we can replace $\partial/\partial r$ with 
$-\partial/\partial t$ in Eq.~(\ref{eq:chireg}). 
However the observed quantities do not depend on the choice 
of the regularization method. This is because the regularization 
is equivalent to 
adding quadratic terms in the first order gauge 
invariant function $\psi^{(1)}_{2\pm 2}$ 
to the second order gauge invariant function $\zeta^{(2)}_{4 \pm 4}$, 
so that it preserves the gauge invariance~\cite{garat00}. 

The explicit expression for the regularized second order source term 
is given by 
\begin{eqnarray}
S_{4 \pm 4}^{reg}(t,r) &=&
\frac{r-2\,M}{42}\, \frac{\sqrt{70}}{\sqrt {\pi }}
\Biggl\{ 
-{\frac {1}{{r}^{6} \left( 3\,r+M \right) ^{2} 
\left( 2\,r+3\,M \right) ^{3}}}
 \left( 72\,{r}^{8}+3936\,{r}^{7}M+2316\,{r}^{6}{M}^{2}- 2030\,{r}^{5}
{M}^{3}
\right.
\nonumber\\
&& 
\left. 
 - 7744\,{r}^{4}{M}^{4} -9512\,{r}^{3}{M}^{5} - 3540\,{r}^{2}{M}^{6}
-1119\,r{M}^{7}-144\,{M}^{8} \right) 
\left(
{\frac {\partial }{\partial r}}\psi^{(1)}_{2\pm 2} \left( t,r \right) 
\right)
{\frac {\partial }{\partial t}}\psi^{(1)}_{2\pm 2} \left( t,r \right) 
\nonumber\\
&& 
+ 
{\frac {1}{{r}^{3} \left( 3\,r+M \right) ^{2} 
\left( 2\,r+3\,M \right) ^{2} \left( r-2\,M \right) ^{2}}}
\left( 24\,{r}^{7} +344\,{r}^{6}M -872\,{r}^{5}{M}^{2}
- 771\,{r}^{4}{M}^{3} 
\right.
\nonumber\\
&& \left. +120\,{r}^{3}{M}^{4}+77\,{r}^{2}{M}^{5}-237\,r{M}^{6}- 48\,{M}^{7}
\right)
\left(
{\frac {\partial ^{2}}{\partial t^2}}\psi^{(1)}_{2\pm 2} \left( t,r \right) 
\right)
{\frac {\partial }{\partial t}}\psi^{(1)}_{2\pm 2} \left( t,r \right) 
\nonumber\\
&& 
-{\frac { 66\,{r}^{4}-106\,{r}^{3}M-220\,{r}^{2}{M}^{2}
-156\,r{M}^{3}-45\,{M}^{4}  }{r \left( M+3\,r
 \right) ^{2} \left( 2\,r+3\,M \right)  \left( r-2\,M \right) }}
\left(
{\frac {\partial ^{3}}{\partial r\partial t^2}}\psi^{(1)}_{2\pm 2} \left( t,r \right) 
\right)
{\frac {\partial }{\partial t}}\psi^{(1)}_{2\pm 2} \left( t,r \right) 
\nonumber\\
&& 
- \frac {3}{{r}^{7} \left( 2\,r+3\,M \right) ^{4}
 \left( 3\,r+M \right) ^{2}}
\left( 2160\,{r}^{9}+11760\,{r}^{8}M
+30560\,{r}^{7}{M}^{2}+41124\,{r}^{6}{M}^{3}+31596\,{r}^{5}{M}^{4}
\right.
\nonumber\\
&&  \left. 
+11630\,{r}^{4}{M}^{5}
-1296\,{r}^{3}{M}^{6}-4182\,{r}^{2}{M}^{7}-1341\,r{M}^{8}-144\,{M}^{9} \right)
\psi^{(1)}_{2\pm 2} \left( t,r \right) 
{\frac {\partial }{\partial t}}\psi^{(1)}_{2\pm 2} \left( t,r \right) 
\nonumber\\
&& 
+ {\frac { 1 }{{r}^{5} \left( 3\,r+M \right) ^{2} 
\left( 2\,r+3\,M \right) ^{2}}}
\left( 228\,{r}^{7} + 8\,{r}^{6}M
-370\,{r}^{5}{M}^{2}+142\,{r}^{4}{M}^{3}-384\,{r}^{3}{M}^{4}
\right.
\nonumber\\
&& 
- \left. 514\,{r}^{2}{M}^{5}-273\,r{M}^{6}-48\,{M}^{7} \right) 
\left(
{\frac {\partial ^{2}}{\partial r\partial t}}\psi^{(1)}_{2\pm 2} \left( t,r \right) 
\right)
{\frac {\partial }{\partial r}}\psi^{(1)}_{2\pm 2} \left( t,r \right) 
\nonumber\\
&& 
+ {\frac { \left( 198\,{r}^{5}-318\,{r}^{4}M -664\,{r}^{3}{M}^{2}
-458\,{r}^{2}{M}^{3}-127\,r{M}^{4}-24\,{M}^{5} \right) }
{3\,{r}^{2} \left( 3\,r+M \right) ^{2}
 \left( 2\,r+3\,M \right)  \left( r-2\,M \right) }} 
\left(
{\frac {\partial ^{3}}{\partial t^3}}\psi^{(1)}_{2\pm 2} \left( t,r \right) 
\right)
{\frac {\partial }{\partial r}}\psi^{(1)}_{2\pm 2} \left( t,r \right) 
\nonumber\\
&& 
- {\frac {
 2 \left( r-2\,M \right) {M}^{2} }{3 \left( 3\,r+M \right) ^{2}{r}^{2}}}
\left(
{\frac {\partial ^{2}}{\partial t^2}}\psi^{(1)}_{2\pm 2} \left( t,r \right) 
\right)
{\frac {\partial ^{2}}{\partial r\partial t}}\psi^{(1)}_{2\pm 2} \left( t,r \right) 
-{\frac { 7\,r+4\,M  }{3\,r}}
\left(
{\frac {\partial ^{3}}{\partial r\partial t^2}}\psi^{(1)}_{2\pm 2} \left( t,r \right) 
\right)
{\frac {\partial ^{2}}{\partial r\partial t}}\psi^{(1)}_{2\pm 2} \left( t,r \right) 
\nonumber\\
&& 
- {\frac {1}{{r}^{6} \left( 3\,r+M \right) ^{2} \left( 2\,r+3\,M \right) ^{3}}} 
\left( 216\,{r}^{8}+4296\,{r}^{7}M
+1992\,{r}^{6}{M}^{2}-3488\,{r}^{5}{M}^{3}- 8716\,{r}^{4}{M}^{4}
\right.
\nonumber\\
&& \left.
-9512\,{r}^{3}{M}^{5}-3540\,{r}^{2}{M}^{6}
-1119\,r{M}^{7}-144\,{M}^{8} \right)
\psi^{(1)}_{2\pm 2} \left( t,r \right) 
{\frac {\partial ^{2}}{\partial r\partial t}}\psi^{(1)}_{2\pm 2} \left( t,r \right) 
\nonumber\\
&& 
+ {\frac { 1 }{{r}^{3} \left( 3\,r+M \right) ^{2}
 \left( 2\,r+3\,M \right) ^{2} \left( r-2\,M \right) }} 
\left( 252\,{r}^{6} +636\,{r}^{5}M+674\,{r}^{4}{M}^{2}
+730\,{r}^{3}{M}^{3}+524\,{r}^{2}{M}^{4}
\right.
\nonumber\\
&& \left.
+171\,r{M}^{5}+24\,{M}^{6}\right)
\psi^{(1)}_{2\pm 2} \left( t,r \right) 
{\frac {\partial ^{3}}{\partial t^3}}\psi^{(1)}_{2\pm 2} \left( t,r \right) 
+ {\frac { \left( 7\,r+4\,M \right) r  }{3 \left( r-2\,M \right) ^{2}}}
\left(
{\frac {\partial ^{3}}{\partial t^3}}\psi^{(1)}_{2\pm 2} \left( t,r \right) 
\right)
{\frac {\partial ^{2}}{\partial t^2}}\psi^{(1)}_{2\pm 2} \left( t,r \right) 
\Biggr\} \,. 
\label{eq:Sreg}
\end{eqnarray}
In the above equation, we note that the second order source term 
is quadratic in the first order wave-function 
$\psi^{(1)}_{2\pm 2}$ 
and hence the second order QNMs have a frequency at 
$\omega^{(2)}_{44}=2\omega^{(1)}_{22}$. 
Since the second order frequencies are 
different from the first order ones, 
we can in principle identify gravitational waves 
from the second order QNMs.

\section{Gauge transformation to asymptotic flat gauge}\label{sec:AF}

In this section, 
in order to obtain physical information from 
the wave-functions, $\psi^{(1)}_{2\pm 2}$ 
and $\chi^{(2)\,reg}_{4 \pm 4}$, 
we treat a gauge transformation 
in the first and second order calculation. 
To obtain the gravitational waveform, it is necessary to go 
to an asymptotic flat (AF) gauge from the RW gauge. 
This is because the Regge-Wheeler-Zerilli formalism 
that we have employed is under the RW gauge 
and this gauge is not asymptotically flat. 

Here, we consider the following gauge transformation~\cite{Mukhanov:1996ak,Bruni:1996im}, 
\begin{eqnarray}
x^{\mu}_{RW} &\to& x^{\mu}_{AF} 
= x^{\mu}_{RW} + \xi^{(1)\mu} \left(x^{\alpha}\right)
+\frac{1}{2} \left[\xi^{(2)\mu}\left(x^{\alpha}\right)
+\xi^{(1)\nu} \xi^{(1)\mu}{}_{,\nu}\left(x^{\alpha}\right)\right] \,,
\end{eqnarray}
where 
comma "," in the index indicates the partial derivative with respect to 
the background coordinates, 
and $\xi^{(1)\mu}$ and $\xi^{(2)\mu}$ are generators of 
the first and second order gauge transformation, respectively. 
By using this gauge transformation, the metric perturbation 
changes as 
\begin{eqnarray}
h_{RW \mu \nu}^{(1)} &\to& h_{AF \mu \nu}^{(1)}=
h_{RW \mu \nu}^{(1)} - {\cal L}_{\xi^{(1)}} g_{\mu \nu} \,,
\label{eq:genGT1}
\\
h_{RW \mu \nu}^{(2)} &\to& h_{AF \mu \nu}^{(2)}=
h_{RW \mu \nu}^{(2)} 
-\frac{1}{2} {\cal L}_{\xi^{(2)}} g_{\mu \nu}
+\frac{1}{2} {\cal L}_{\xi^{(1)}}^2 g_{\mu \nu}
-{\cal L}_{\xi^{(1)}} h_{RW \mu \nu}^{(1)} \,,
\label{eq:genGT2}
\end{eqnarray}
where ${\cal L}_{\xi^{(i)}}$ means the Lie derivative 
in the $\xi^{(i)}$ direction. 
We use the following form of a generator for the gauge transformation, 
\begin{eqnarray}
\xi^{(i)\mu} &=& 
\{V_0^{(i)}(t,r) Y_{\ell m}(\theta,\phi),\,V_1^{(i)}(t,r) Y_{\ell m}(\theta,\phi),\,
V_2^{(i)}(t,r) \frac{\partial}{\partial \theta} Y_{\ell m}(\theta,\phi),\,
V_2^{(i)}(t,r) \frac{1}{\sin^2 \theta} 
\frac{\partial}{\partial \phi} Y_{\ell m}(\theta,\phi)\} \,,
\label{eq:GoGT}
\end{eqnarray}
where we have considered only the even parity mode, 
and hence we have three degrees of gauge freedom for each order. 
(Note that the generator for the odd parity part is 
$\xi^{(i)\mu}=\{0,\,0,\,-V_3^{(i)} \partial_{\phi}Y_{\ell m}/\sin \theta,\,
V_3^{(i)} \sin \theta \,\partial_{\theta} Y_{\ell m}\}$.) 
In this paper, 
we consider only the $\ell=2,\,m=\pm2$ modes for the first order gauge transformation, 
and the $\ell=4,\,m=\pm4$ modes for the second order one. 

The gauge transformation of the metric perturbation is explicitly given as follows. 
For the first order metric perturbation, we derive 
\begin{eqnarray}
H_{0\,2\pm2}^{(1)AF}(t,r) &=& H_{0\,2\pm2}^{(1)RW}(t,r)
+ 2\,{\frac {\partial }{\partial t}}{V_0}^{(1)} \left( t,r \right) +2\,{
\frac {M  }{r \left( r-2\,M \right) }}{V_1}^{(1)}\left( t,r \right) \,, 
\nonumber \\ 
H_{1\,2\pm2}^{(1)AF}(t,r) &=& H_{1\,2\pm2}^{(1)RW}(t,r)
+ {\frac { \left( r-2\,M \right)  }{r}}{\frac {\partial }{\partial r}}{V_0}^{(1)}
 \left( t,r \right)
-{\frac {r}{r-2\,M}}{\frac {\partial }{\partial t}}{
V_1}^{(1)} \left( t,r \right) \,,
\nonumber \\ 
H_{2\,2\pm2}^{(1)AF}(t,r) &=& H_{2\,2\pm2}^{(1)RW}(t,r)
-2\,{\frac {\partial }{\partial r}}{V_1}^{(1)} \left( t,r \right) +2\,{
\frac {M}{r \left( r-2\,M \right) }} {V_1}^{(1)} \left( t,r \right) \,,
\nonumber \\ 
K_{2\pm2}^{(1)AF}(t,r) &=& K_{0\,2\pm2}^{(1)RW}(t,r)
-2\,{\frac {1}{r}} {V_1}^{(1)} \left( t,r \right) \,,
\nonumber \\ 
h_{0\,2\pm2}^{(1)(e)AF}(t,r) &=& 
{\frac { \left( r-2\,M \right) }{r}}{V_0}^{(1)} \left( t,r \right) -{r}^{
2}{\frac {\partial }{\partial t}}{V_2}^{(1)} \left( t,r \right) \,,
\nonumber \\ 
h_{1\,2\pm2}^{(1)(e)AF}(t,r) &=& 
-{\frac {r }{r-2\,M}}{V_1}^{(1)} \left( t,r \right)-{r}^{2}{\frac {
\partial }{\partial r}}{V_2}^{(1)} \left( t,r \right) \,,
\nonumber \\ 
G_{2\pm2}^{(1)AF}(t,r) &=& 
-2\,{V_2}^{(1)} \left( t,r \right) \,.
\end{eqnarray}
For the second order metric perturbation, 
we can calculate the gauge transformation straightforward, 
but obtain very long expressions. For example, 
they are written formally as 
\begin{eqnarray}
H_{0\,4\pm4}^{(2)AF}(t,r) &=& H_{0\,4\pm4}^{(2)RW}(t,r)
+ {\frac {\partial }{\partial t}}{V_0}^{(2)} \left( t,r \right) 
+ {\frac {M}{r \left( r-2\,M \right) }}{V_1}^{(2)} \left( t,r \right) 
+\delta H^{(2)}_{0\,4 \pm 4} \left( t,r \right) \,, 
\\
K_{4 \pm 4}^{(2)AF}(t,r) &=& K_{4 \pm 4}^{(2)RW}(t,r) 
-{\frac {1 }{r}}{V_1}^{(2)} \left( t,r \right)
+\delta K^{(2)}_{4 \pm 4} \left( t,r \right) \,,
\label{eq:2ndKGT-gen} \\
h_{1\,4 \pm 4}^{(2)(e)AF}(t,r) &=& 
-{\frac {r}{2\,(r-2\,M)}}{V_1}^{(2)} \left( t,r \right) -\frac{{r}^{2}}{2}
{\frac {\partial }{\partial r}}{V_2}^{(2)} \left( t,r \right) 
+ \delta h_{1\,4 \pm 4}^{(2)(e)} \left( t,r \right) \,.
\label{eq:2ndh1eGT-gen} \\
G_{4 \pm 4}^{(2)AF}(t,r) &=& 
-{V_2}^{(2)} \left( t,r \right) +\delta G^{(2)}_{4 \pm 4} \left( t,r \right)
\,,
\label{eq:2ndGT-gen}
\end{eqnarray}
where $\delta H^{(2)}_{0\,4 \pm 4}$, $\delta K^{(2)}_{4 \pm 4}$, 
$\delta h_{1\,4 \pm 4}^{(2)(e)}$ and $\delta G^{(2)}_{4 \pm 4}$ 
are defined by the tensor harmonics expansion of 
the last two terms in the right hand side 
of Eq.~(\ref{eq:genGT2}), i.e., 
$(1/2) {\cal L}_{\xi^{(1)}}^2 g_{\mu \nu}^{(0)}-{\cal L}_{\xi^{(1)}} h_{RW \mu \nu}^{(1)}$. 
This includes only quadratic terms of 
the first order wave-function because 
$h_{RW \mu \nu}^{(1)}$ and $\xi^{(1)\mu}$ 
are the first order quantities written by 
the first order wave-function $\psi^{(1)}_{2\pm 2}$
after we solve the gauge equations.

\subsection{About first order}

First, we consider the asymptotic behavior 
of the metric perturbation at large $r$ in the RW gauge. 
Using Eq.~(\ref{eq:1stREC}) and 
the asymptotic expansion in Eq.~(\ref{eq:asympt1st}), 
the metric perturbation is given as follows,
\begin{eqnarray}
H_{0\,2\pm2}^{(1)RW}(t,r) &=& H_{2\,2\pm2}^{(1)RW}(t,r) 
\nonumber \\ &=& 
\frac{1}{3}\, \left( {\frac {d^{4}}{d{{\it T_r}}^{4}}}F_I \left( {\it T_r}
 \right)  \right) r+\frac{2}{3}\, \left( {\frac {d^{4}}{d{{\it T_r}}^{4}}}F_I
 \left( {\it T_r} \right)  \right) M+\frac{2}{3}\,{\frac {d^{3}}{d{{\it T_r}}^{3
}}}F_I \left( {\it T_r} \right) 
\nonumber \\ &&
+ \left( {\frac {d^{2}}{d{{\it T_r}}^{2}}}
F_I \left( {\it T_r} \right) +{\frac {11}{6}}\, \left( {\frac {d^{3}}{d{{
\it T_r}}^{3}}}F_I \left( {\it T_r} \right)  \right) M+\frac{4}{3}\, \left( {
\frac {d^{4}}{d{{\it T_r}}^{4}}}F_I \left( {\it T_r} \right)  \right) {M}^
{2} \right) \frac{1}{r} + O(r^{-2}) \,,
\nonumber \\ 
H_{1\,2\pm2}^{(1)RW}(t,r) &=& 
-\frac{1}{3}\, \left( {\frac {d^{4}}{d{{\it T_r}}^{4}}}F_I \left( {\it T_r}
 \right)  \right) r-\frac{2}{3}\, \left( {\frac {d^{4}}{d{{\it T_r}}^{4}}}F_I
 \left( {\it T_r} \right)  \right) M-\frac{2}{3}\,{\frac {d^{3}}{d{{\it T_r}}^{3
}}}F_I \left( {\it T_r} \right) 
\nonumber \\  
&& 
+ \left( -\frac{4}{3}\, \left( {\frac {d^{4}}{d{{
\it T_r}}^{4}}}F_I \left( {\it T_r} \right)  \right) {M}^{2}-{\frac {d^{2}
}{d{{\it T_r}}^{2}}}F_I \left( {\it T_r} \right) -{\frac {11}{6}}\,
 \left( {\frac {d^{3}}{d{{\it T_r}}^{3}}}F_I \left( {\it T_r} \right) 
 \right) M \right) \frac{1}{r} +O(r^{-2}) \,,
\nonumber \\ 
K_{2\pm2}^{(1)RW}(t,r) &=& 
-\frac{1}{3}\,{\frac {d^{3}}{d{{\it T_r}}^{3}}}F_I \left( {\it T_r} \right) 
+{
\frac {1 }{{r}^{2}}}
\left(\frac{1}{2}\, \left( {\frac {d^{2}}{d{{\it T_r}}^{2}}}F_I \left( {\it T_r}
 \right)  \right) M+{\frac {d}{d{\it T_r}}}F_I \left( {\it T_r} \right)\right) 
+ O(r^{-3}) 
\,,
\label{eq:1stRWA}
\end{eqnarray}
where we introduce ${\it T_r} = t-r_*(r)$ for simplicity. 
Since we are interested in the asymptotic behavior now 
and the $m=2$ and $m=-2$ modes have the same asymptotic behavior, 
we have used the same notation for the $m=2$ and $m=-2$ modes. 

On the other hand, the metric perturbation in an AF gauge 
should behave as 
\begin{eqnarray}
H_{0\,2\pm2}^{(1)AF}(t,r) &=& H_{1\,2\pm2}^{(1)AF}(t,r)
=h_{0\,2\pm2}^{(1)(e)AF}(t,r)=0 \,,
\nonumber \\ 
H_{2\,2\pm2}^{(1)AF}(t,r) &=& O(r^{-3}) \,,
\nonumber \\ 
h_{1\,2\pm2}^{(1)(e)AF}(t,r) &=& O(r^{-1}) \,,
\nonumber \\ 
K_{2\pm2}^{(1)AF}(t,r) &=& O(r^{-1}) \,,
\nonumber \\ 
G_{2\pm2}^{(1)AF}(t,r) &=& O(r^{-1}) \,.
\label{eq:AFgauge}
\end{eqnarray}
This asymptotic behavior will be also used 
for the second order calculation. 
Then, we can find that the gauge transformation has the following form. 
\begin{eqnarray}
V_0^{(1)}(t,r) &=& 
-\frac{1}{6}\, \left( {\frac {d^{3}}{d{{\it T_r}}^{3}}}F_I \left( {\it T_r}
 \right)  \right) r-\frac{1}{3}\,{\frac {d^{2}}{d{{\it T_r}}^{2}}}F_I \left( {
\it T_r} \right) 
-\frac{1}{3}\, \left( {\frac {d^{3}}{d{{\it T_r}}^{3}}}F_I
 \left( {\it T_r} \right)  \right) M
\nonumber \\ && 
+{\frac {1}{r}}
\left(
-\frac{3}{4}\, \left( {\frac {d^{2
}}{d{{\it T_r}}^{2}}}F_I \left( {\it T_r} \right)  \right) M-\frac{2}{3}\, \left( 
{\frac {d^{3}}{d{{\it T_r}}^{3}}}F_I \left( {\it T_r} \right)  \right) {M}
^{2}-\frac{1}{2}\,{\frac {d}{d{\it T_r}}}F_I \left( {\it T_r} \right) 
\right)  
+ O(r^{-2}) \,, 
\nonumber \\
V_1^{(1)}(t,r) &=& 
-\frac{1}{6}\, \left( {\frac {d^{3}}{d{{\it T_r}}^{3}}}F_I \left( {\it T_r}
 \right)  \right) r-\frac{1}{2}\,{\frac {d^{2}}{d{{\it T_r}}^{2}}}F_I \left( {
\it T_r} \right) 
+{\frac {1}{r}}
\left(
-\frac{1}{2}\,{\frac {d}{d{\it T_r}}}F_I \left( {\it T_r}
 \right) +\frac{1}{4}\, \left( {\frac {d^{2}}{d{{\it T_r}}^{2}}}F_I \left( {\it 
T_r} \right)  \right) M
\right) + O(r^{-2}) \,,
\nonumber \\ 
V_2^{(1)}(t,r) &=& 
-\frac{1}{6}\,{\frac {1}{r}}{\frac {d^{2}}{d{{\it T_r}}^{2}}}F_I \left( {\it T_r}
 \right) 
-\frac{1}{3}\,{\frac {1}{{r}^{2}}}
{\frac {d}{d{\it T_r}}}F_I \left( {\it T_r}
 \right) 
+{\frac {1}{{r}^{3}}}
\left(
-\frac{1}{12}\,M{\frac {d}{d{\it T_r}}}F_I \left( {
\it T_r} \right) -\frac{1}{2}\,F_I \left( {\it T_r} \right) 
\right) 
+ O(r^{-4}) 
\,.
\label{eq:1stGT}
\end{eqnarray}
The above results are obtained iteratively for the large $r$ expansion. 
Here, we note that the transverse-traceless tensor harmonics 
for the even parity part is $\bm{f}_{\ell m}$ 
in Eqs.~(\ref{eq:hharm}) and (\ref{eq:flm}). 
Therefore the coefficient of the metric perturbation 
that is related to the gravitational wave is
$G_{\ell m}^{(1)AF}$. 
With Eqs.~(\ref{eq:2ndGT-gen}) and (\ref{eq:1stGT}), we obtain 
\begin{eqnarray}
G_{2\pm2}^{(1)AF}(t,r) &=& 
\frac{1}{3}\,{\frac {1}{r}}{\frac {d^{2}}{d{{\it T_r}}^{2}}}F_I \left( {\it T_r}
 \right) +O(r^{-2}) \,.
\end{eqnarray}
This can be shown in terms of $\psi^{(1)}_{2\pm 2}$ 
with Eq.~(\ref{eq:asympt1st}) as 
\begin{eqnarray}
G_{2\pm2}^{(1)AF}(t,r) &=& 
{\frac {1}{r}}\psi^{(1)}_{2\pm 2}(t,r) 
+O(r^{-2}) \,. 
\label{eq:1stwave}
\end{eqnarray}

\subsection{About second order}

In order to derive the gravitational wave amplitude 
for the second perturbative order, 
we also need to obtain the coefficient $G_{4 \pm 4}^{(2)AF}$ 
under the AF gauge as in the first order case. 
From Eq.~(\ref{eq:2ndGT-gen}), 
it is found that we must derive $\delta G^{(2)}_{4 \pm 4}$ and 
${V_2}^{(2)}$ which is calculated by using the solution of ${V_1}^{(2)}$. 
In the following, we first obtain ${V_1}^{(2)}$ as Eq.~(\ref{eq:V21}) 
by solving the gauge equation in Eq.~(\ref{eq:2ndKGT-gen}). 
Then, ${V_2}^{(2)}$ is derived from Eq.~(\ref{eq:2ndh1eGT-gen}) 
as in Eq.~(\ref{eq:V22}) 
and $\delta G^{(2)}_{4 \pm 4}$ is calculated as in Eq.~(\ref{eq:delG}). 

First we consider to obtain $V_1^{(2)}$ in terms of the asymptotic wave-functions. 
The equation 
\begin{eqnarray}
K_{4 \pm 4}^{(2)AF}(t,r) &=& K_{4 \pm 4}^{(2)RW}(t,r) 
-{\frac {1}{r}}{V_1}^{(2)} \left( t,r \right) 
+\delta K^{(2)}_{4 \pm 4}(t,r) \,,
\label{eq:2ndKGT}
\end{eqnarray}
is used as shown in Eq.~(\ref{eq:2ndKGT-gen}). 
Here $K_{4 \pm 4}^{(2)RW}$ is derived from 
$\chi^{(2)}_{4 \pm 4}$, ${\cal A}_{1\,4 \pm 4}$
and ${\cal B}_{0\,4 \pm 4}$ 
in Eqs.~(\ref{eq:rec2nd}).
Since the wave-function $\chi^{(2)}_{4 \pm 4}$
is given by 
$\chi^{(2)\,reg}_{4 \pm 4}$ and $\psi^{(1)}_{2 \pm 2}$
in Eq.~(\ref{eq:chireg}) 
and the coefficients ${\cal A}_{1\,4 \pm 4}$
and ${\cal B}_{0\,4 \pm 4}$ 
are also given by $\psi^{(1)}_{2 \pm 2}$ 
with Eqs.~(\ref{eq:calT}), (\ref{eq:Tharm}) and (\ref{eq:1stREC}),
we may obtain $K_{4 \pm 4}^{(2)RW}$
in terms of wave-functions 
$\chi^{(2)\,reg}_{4 \pm 4}$ and $\psi^{(1)}_{2 \pm 2}$.
The asymptotic expansion of $\psi^{(1)}_{2 \pm 2}$ is 
given by Eq.~(\ref{eq:asympt1st}) 
while the asymptotic expansion of $\chi^{(2)\,reg}_{4 \pm 4}$ 
is obtained from the Zerilli equation (\ref{eq:Zeqchi-reg}) as 
\begin{eqnarray}
\chi^{(2)\,reg}_{4 \pm 4}(t,r) &=& 
F_I^{(2)} \left( {\it T_r} \right) + O(r^{-1})  \,,
\end{eqnarray}
where $F_I^{(2)}$ is some second order function. 
Here we consider only the leading behavior for large $r$, 
because this is sufficient in order to obtain the final gravitational waveform. 
If we consider the higher order calculation, 
it should note that there is a contribution from 
the second order source term of Eq.~(\ref{eq:Sreg}) in the above equation. 
Then, we can calculate the asymptotic expansion of
$\partial K_{4 \pm 4}^{(2)RW} /\partial t$ in Eqs.~(\ref{eq:rec2nd}) as 
\begin{eqnarray}
{\frac {\partial }{\partial t}}K_{4 \pm 4}^{(2)RW} \left( t,r \right) 
&=& 
- {\frac {d}{d{{\it T_r}}}}F_I^{(2)} \left( {\it T_r} \right) 
\nonumber \\ && 
-{\frac {\sqrt {70}}{3402\sqrt {\pi }}}
\Biggl(  
\left( 
 \left( {\frac {d^{5}}{d{{\it T_r}}^{5}}}F_I \left( {\it T_r} \right) 
 \right) {\frac {d^{3}}{d{{\it T_r}}^{3}}}F_I \left( {\it T_r} \right) +
 \left( {\frac {d^{4}}{d{{\it T_r}}^{4}}}F_I \left( {\it T_r} \right) 
 \right) ^{2} \right) M
\nonumber \\ && \quad 
+9\, \left( {\frac {d^{5}}{d{{\it T_r}}^{5}}}F_I
 \left( {\it T_r} \right)  \right) {\frac {d^{2}}{d{{\it T_r}}^{2}}}F_I
 \left( {\it T_r} \right) -15\, \left( {\frac {d^{4}}{d{{\it T_r}}^{4}}}
F_I \left( {\it T_r} \right)  \right) {\frac {d^{3}}{d{{\it T_r}}^{3}}}F_I
 \left( {\it T_r} \right) \Biggr) + O(r^{-1}) \,. 
\end{eqnarray}
Integrating the above equation with respect to $t$, 
the second order metric component $K_{4 \pm 4}^{(2)RW}$ becomes 
\begin{eqnarray}
K_{4 \pm 4}^{(2)RW} \left( t,r \right) 
&=& 
- F_I^{(2)} \left( {\it T_r} \right) 
\nonumber \\ && 
-{\frac {\sqrt{70}}{3402\sqrt{\pi}}}\,
\Biggl( 9\,
 \left( {\frac {d^{4}}{d{{\it T_r}}^{4}}}F_I \left( {\it T_r} \right) 
 \right) {\frac {d^{2}}{d{{\it T_r}}^{2}}}F_I \left( {\it T_r} \right) -12
\, \left( {\frac {d^{3}}{d{{\it T_r}}^{3}}}F_I \left( {\it T_r} \right) 
 \right) ^{2}
\nonumber \\ && 
+ \left( {\frac {d^{3}}{d{{\it T_r}}^{3}}}F_I \left( {\it T_r
} \right)  \right)  \left( {\frac {d^{4}}{d{{\it T_r}}^{4}}}F_I \left( {
\it T_r} \right)  \right) M \Biggr) + O(r^{-1}) 
 \,.
\end{eqnarray}
In Eq.~(\ref{eq:2ndKGT}), 
we also need the asymptotic expansion of 
$\delta K^{(2)}_{4 \pm 4}$, 
which is defined by the tensor harmonics expansion of
$(1/2) {\cal L}_{\xi^{(1)}}^2 g_{\mu \nu}^{(0)}
-{\cal L}_{\xi^{(1)}} h_{RW \mu \nu}^{(1)}$ in Eq.~(\ref{eq:genGT2}). 
Using the first order metric perturbation under the RW gauge 
in Eq.~(\ref{eq:1stRWA}) and the generator 
of the first order gauge transformation in Eq.~(\ref{eq:GoGT}) 
with Eq.~(\ref{eq:1stGT}), 
$\delta K^{(2)}_{4 \pm 4}$ is derived as 
\begin{eqnarray}
\delta K^{(2)}_{4 \pm 4} 
&=& {\frac {\sqrt {70} }{504\sqrt {\pi }}}\,
\left(  \left( {\frac {d^{4}}{d{{
\it T_r}}^{4}}}F_I \left( {\it T_r} \right)  \right) {\frac {d^{2}}{d{{
\it T_r}}^{2}}}F_I \left( {\it T_r} \right) -2\, \left( {\frac {d^{3}}{d{{
\it T_r}}^{3}}}F_I \left( {\it T_r} \right)  \right) ^{2} \right)
 + O(r^{-1}) \,,
\end{eqnarray}
From the above results and the AF gauge condition of $K^{(2)}_{4 \pm 4}$ 
in Eq.~(\ref{eq:AFgauge}), 
in order to remove the $O(r^0)$ terms from $K^{(2)RW}_{4 \pm 4}$ 
and $\delta K^{(2)}_{4 \pm 4}$ in Eq.~(\ref{eq:2ndKGT}), 
we obtain the pure second order gauge transformation ${V_1}^{(2)}$ as 
\begin{eqnarray}
{V_1}^{(2)} \left( t,r \right) &=& 
-r\, F_I^{(2)} \left( {\it T_r} \right) 
\nonumber \\ && 
-{\frac {\sqrt{70}}{13608\sqrt {\pi }}}\,r
\Biggl( 9\,
 \left( {\frac {d^{4}}{d{{\it T_r}}^{4}}}F_I \left( {\it T_r} \right) 
 \right) {\frac {d^{2}}{d{{\it T_r}}^{2}}}F_I \left( {\it T_r} \right) +6
\, \left( {\frac {d^{3}}{d{{\it T_r}}^{3}}}F_I \left( {\it T_r} \right) 
 \right) ^{2}
\nonumber \\ && \quad 
+4\, \left( {\frac {d^{3}}{d{{\it T_r}}^{3}}}F_I \left( {
\it T_r} \right)  \right)  \left( {\frac {d^{4}}{d{{\it T_r}}^{4}}}F_I
 \left( {\it T_r} \right)  \right) M 
\Biggr) + O(r^0)  \,.
\label{eq:V21}
\end{eqnarray}

Next, we derive ${V_2}^{(2)}$ from the AF gauge condition 
of $h_{1\,4 \pm 4}^{(2)(e)}$ in Eq.~(\ref{eq:AFgauge}). 
The second order gauge transformation of this component is given by 
\begin{eqnarray}
h_{1\,4 \pm 4}^{(2)(e)AF}(t,r) &=& 
-{\frac {r }{2\,(r-2\,M)}}{V_1}^{(2)} \left( t,r \right)-\frac{{r}^{2}}{2}
{\frac {\partial }{\partial r}}{V_2}^{(2)} \left( t,r \right) 
+ \delta h_{1\,4 \pm 4}^{(2)(e)}  \left( t,r \right) \,,
\label{eq:2ndh1eGT}
\end{eqnarray}
as shown in Eq.~(\ref{eq:2ndh1eGT-gen}). 
Here we have used the fact that $h_{1\,4 \pm 4}^{(2)(e)RW}=0$ under the RW gauge. 
We may also derive
$\delta h_{1\,4 \pm 4}^{(2)(e)}$, which is defined
by the tensor harmonics expansion of
$(1/2) {\cal L}_{\xi^{(1)}}^2 g_{\mu \nu}^{(0)}
-{\cal L}_{\xi^{(1)}} h_{RW \mu \nu}^{(1)}$ in Eq.~(\ref{eq:genGT2}), 
by Eqs.~(\ref{eq:1stRWA}) 
and (\ref{eq:GoGT}) with Eq.~(\ref{eq:1stGT}) as 
\begin{eqnarray}
\delta h_{1\,4 \pm 4}^{(2)(e)} &=& 
{\frac {\sqrt {70}}{1008\sqrt {\pi }}}\,r\,
\left(  \left( {\frac {d^{4}}{d{
{\it T_r}}^{4}}}F_I \left( {\it T_r} \right)  \right) {\frac {d^{2}}{d{{
\it T_r}}^{2}}}F_I \left( {\it T_r} \right) + \left( {\frac {d^{3}}{d{{
\it T_r}}^{3}}}F_I \left( {\it T_r} \right)  \right) ^{2} \right) 
 + O(r^0) 
\,.
\end{eqnarray}
Then, ${V_2}^{(2)}$ is calculated from the above value and the result 
of ${V_1}^{(2)}$ in Eq.~(\ref{eq:V21}) 
with the AF gauge condition in Eq.(\ref{eq:AFgauge}) as 
\begin{eqnarray}
{\frac {\partial }{\partial r}}{V_2}^{(2)} \left( t,r \right) 
&=& 
\frac{1}{r}\,F_I^{(2)} \left( {\it T_r} \right) 
\nonumber \\ && 
+ {\frac {\sqrt{70}}{13608\sqrt {\pi }}}
\,\frac{1}{r}
\,\Biggl( 36\,
 \left( {\frac {d^{4}}{d{{\it T_r}}^{4}}}F_I \left( {\it T_r} \right) 
 \right) {\frac {d^{2}}{d{{\it T_r}}^{2}}}F_I \left( {\it T_r} \right) 
+33\, \left( {\frac {d^{3}}{d{{\it T_r}}^{3}}}F_I
 \left( {\it T_r} \right)  \right) ^{2} 
\nonumber \\ && \quad 
+4
\, \left( {\frac {d^{3}}{d{{\it T_r}}^{3}}}F_I \left( {\it T_r} \right) 
 \right)  \left( {\frac {d^{4}}{d{{\it T_r}}^{4}}}F_I \left( {\it T_r}
 \right)  \right) M\Biggr) 
+ O(r^{-2}) 
\nonumber \\ 
&=& - {\frac {\partial }{\partial t}}{V_2}^{(2)} \left( t,r \right) 
+ O(r^{-2})  \,.
\label{eq:V22}
\end{eqnarray}
In the last line of the above equation, 
we have used the definition of $T_r=t-r_*(r)$. 

At this stage, we can consider the metric perturbation related to 
the gravitational wave amplitude, i.e., $G_{4 \pm 4}^{(2)AF}$. 
The gauge transformation of the components $G^{(2)}_{4 \pm 4}$ is given by 
\begin{eqnarray}
G_{4 \pm 4}^{(2)AF}(t,r) &=& 
-{V_2}^{(2)} \left( t,r \right) +\delta G^{(2)}_{4 \pm 4} 
\,,
\label{eq:2ndGTG}
\end{eqnarray}
as shown in Eq.~(\ref{eq:2ndGT-gen}). 
Here $\delta G^{(2)}_{4 \pm 4}$,
which is defined by the tensor harmonics expansion of
$(1/2) {\cal L}_{\xi^{(1)}}^2 g_{\mu \nu}^{(0)}
-{\cal L}_{\xi^{(1)}} h_{RW \mu \nu}^{(1)}$ in Eq.~(\ref{eq:genGT2}), 
is also obtained from Eqs.~(\ref{eq:1stRWA}) 
and (\ref{eq:GoGT}) with Eq.~(\ref{eq:1stGT}) as 
\begin{eqnarray}
\delta G^{(2)}_{4 \pm 4} 
&=&  -{\frac {\sqrt {70}}{1512\sqrt {\pi }}}
\,\frac{1}{r}\,\left( {\frac {d^{3}}{d{{\it T_r}}^{3}}}F_I
 \left( {\it T_r} \right)  \right)  \left( {\frac {d^{2}}{d{{\it T_r}}^{
2}}}F_I \left( {\it T_r} \right)  \right) 
 + O(r^{-2}) 
 \,.
\label{eq:delG}
\end{eqnarray}
Finally, using Eq.~(\ref{eq:V22}) 
for ${V_2}^{(2)}$ in Eq.~(\ref{eq:2ndGTG}), 
we obtain 
\begin{eqnarray}
{\frac {\partial }{\partial t}}G_{4 \pm 4}^{(2)AF}(t,r) &=& 
\frac{1}{r}\,F_I^{(2)} 
\left( {\it T_r} \right) 
\nonumber \\ && 
+ 
{\frac {\sqrt{70}}{13608\sqrt {\pi }}}\,
\frac{1}{r}\,
\Biggl( 27\,
 \left( {\frac {d^{4}}{d{{\it T_r}}^{4}}}F_I \left( {\it T_r} \right) 
 \right) {\frac {d^{2}}{d{{\it T_r}}^{2}}}F_I \left( {\it T_r} \right) 
+24\, \left( {\frac {d^{3}}{d{{\it T_r}}^{3}}}F_I
 \left( {\it T_r} \right)  \right) ^{2}
\nonumber \\ && \quad 
+4\, \left( {\frac {d^{3}}{d{{\it T_r}}^{3}}}F_I \left( {\it T_r} \right) 
 \right)  \left( {\frac {d^{4}}{d{{\it T_r}}^{4}}}F_I \left( {\it T_r}
 \right)  \right) M \Biggr) 
 + O(r^{-2}) 
 \,.
\end{eqnarray}
By using $\psi^{(1)}_{2\pm 2}$ and $\chi^{(2)\,reg}_{4 \pm 4}$, 
the gravitational waveform is obtained as 
\begin{eqnarray}
{\frac {\partial }{\partial t}}G_{4 \pm 4}^{(2)AF}(t,r) &=& 
\frac{1}{r}\chi^{(2)\,reg}_{4 \pm 4}(t,r) 
\nonumber \\ && 
+ 
{\frac {\sqrt{70}}{1512\sqrt {\pi }}}\,
\frac{1}{r}\,
\Biggl(
27\,\psi^{(1)}_{2\pm 2}(t,r) 
{\frac {\partial^{2}}{\partial t^{2}}}\psi^{(1)}_{2\pm 2}(t,r)
+24\,\left( {\frac {\partial}{\partial t}}\psi^{(1)}_{2\pm 2}(t,r)  \right) ^{2}
\nonumber \\ && \quad 
+4\,M\,\left({\frac {\partial}{\partial t}}\psi^{(1)}_{2\pm 2}(t,r)\right) 
{\frac {\partial^{2}}{\partial t^{2}}}\psi^{(1)}_{2\pm 2}(t,r)
\Biggr) + O(r^{-2}) 
\,.
\label{eq:2ndwave}
\end{eqnarray}
It should be noted that 
we can show that all metric 
components satisfy the asymptotic flat gauge condition.

\subsection{Power of Gravitational waves}

In this subsection, we summarize the power of gravitational waves 
which is derived from the gravitational waveform. 
The power $P$ per solid angle is given by~\cite{landau75} as 
\begin{eqnarray}
\frac{dP}{d\Omega} &=& \frac{1}{16\pi r^2}
\left\langle\frac{1}{\sin^2 \theta}\left(\frac{\partial}{\partial t} h_{\theta \phi}\right)^2
+\frac{1}{4}\left(\frac{\partial}{\partial t} h_{\theta \theta}
-\frac{1}{\sin^2 \theta}\frac{\partial}{\partial t} h_{\phi \phi}\right)^2
\right\rangle\,,
\label{eq:dPdO}
\end{eqnarray}
where $\langle \cdots \rangle$ means an averaging over 
a spacetime region which is sufficiently larger 
than the characteristic wavelength of gravitational waves. 
Therefore, for the $\ell=2,\,m=\pm 2$ modes 
of the first order metric perturbation, 
we have the following equation. 
\begin{eqnarray}
\frac{dP^{(1)}}{d\Omega} &=& 
\frac{r^2}{64\pi}
\Biggl\langle\frac{1}{\sin^2 \theta}
\left(\frac{\partial}{\partial t}G_{22}^{(1)AF}(t,r)X_{22}(\theta,\phi)
+\frac{\partial}{\partial t}G_{2-2}^{(1)AF}(t,r)X_{2-2}(\theta,\phi)\right)^2
\nonumber \\ && \qquad
+ \left(\frac{\partial}{\partial t}G_{22}^{(1)AF}(t,r)W_{22}(\theta,\phi)
+\frac{\partial}{\partial t}G_{2-2}^{(1)AF}(t,r)W_{2-2}(\theta,\phi)\right)^2
\Biggr\rangle
\nonumber \\ &=&
\frac{r^2}{64\pi}\Biggl[
2\,\Biggl\langle\left(\frac{\partial}{\partial t}G_{22}^{(1)AF}(t,r)\right)
\left(\frac{\partial}{\partial t}G_{2-2}^{(1)AF}(t,r)\right)\Biggl\rangle
\left(\frac{1}{\sin^2\theta} X_{22}(\theta,\phi) X_{2-2}(\theta,\phi)
+W_{22}(\theta,\phi) W_{2-2}(\theta,\phi)\right)
\nonumber \\ && \qquad 
+\Biggl\langle\left(\frac{\partial}{\partial t}G_{22}^{(1)AF}(t,r)\right)^2\Biggr\rangle
\left(\frac{1}{\sin^2\theta} \left(X_{22}(\theta,\phi)\right)^2 
+\left(W_{22}(\theta,\phi)\right)^2 \right)
\nonumber \\ && \qquad 
+\Biggl\langle\left(\frac{\partial}{\partial t}G_{2-2}^{(1)AF}(t,r)\right)^2\Biggr\rangle
\left(\frac{1}{\sin^2\theta} \left(X_{2-2}(\theta,\phi)\right)^2 
+\left(W_{2-2}(\theta,\phi)\right)^2 \right)
\Biggr]
\nonumber \\ &=& 
\frac{r^2}{64\pi}
\sum_{m=\pm2}\Biggl[
\left|\frac{\partial}{\partial t}G_{2m}^{(1)AF}(t,r)\right|^2
\left( \frac{1}{\sin^2 \theta}
\left|X_{2m}(\theta,\phi)\right|^2 
+ \left|W_{2m}(\theta,\phi)\right|^2 \right) \nonumber \\ && \qquad \qquad 
+\Biggl\langle\left(\frac{\partial}{\partial t}G_{2m}^{(1)AF}(t,r)\right)^2\Biggr\rangle
\left(\frac{1}{\sin^2\theta} \left(X_{2m}(\theta,\phi)\right)^2 
+\left(W_{2m}(\theta,\phi)\right)^2 \right)
\Biggr]
\,,
\label{eq:dpdo}
\end{eqnarray}
in terms of the coefficient of the tensor harmonics in Eq.~(\ref{eq:hharm}). 
In the above equation, 
we use $G_{22}^{(1)AF}=G_{2-2}^{(1)AF\,*}$, and 
the angular functions $X_{\ell m}$ and $W_{\ell m}$ 
are given by Eqs.~(\ref{eq:Xlm}), (\ref{eq:Wlm}) and (\ref{eq:Y22}).
We note that it is meaningless to distinguish the power of $m=\pm 2$ modes
since they appear as the cross term in the second equality of the above equation. 
It is also noted that the averaging 
$\langle (\partial G_{2m}^{(1)AF}/\partial t)^2\rangle$ 
does not vanish in this calculation. 
Integrating the above equation with respect to the angular directions, 
we obtain
\begin{eqnarray}
P^{(1)} &=& \frac{3}{8\pi}\,r^2 \sum_{m=\pm2}
\left|\frac{\partial}{\partial t}G_{2m}^{(1)AF}(t,r)\right|^2
\nonumber \\ &=& 
\frac{3}{8\pi} \sum_{m=\pm2}
\left|\frac{\partial}{\partial t}\psi^{(1)}_{2m}(t,r)\right|^2
\nonumber \\ &=& 
\frac{3}{4\pi} \left|\frac{\partial}{\partial t}\psi^{(1)}_{22}(t,r) \right|^2
 \,, 
\label{eq:GG1}
\end{eqnarray}
where we have used the result in Eq.~(\ref{eq:1stwave}) 
in the second line 
and the fact that $\psi^{(1)}_{2-2}=\psi^{(1)\,*}_{22}$ 
in Eq.~(\ref{eq:1stSET0}) in the last line. 
The relation between
$m=2$ and $m=-2$ modes are summarized in the appendix. 

When we set $\psi^{(1)}_{22} = A\, \exp(-i \omega_{\ell=2,m=2}^{(1)} (t-r_*))$, 
where we choose the amplitude $A$ of $\psi^{(1)}_{22}$ 
at the origin of $t-r_*$, the total radiated energy is 
\begin{eqnarray}
E^{(1)} &=& \int dt \,P^{(1)} 
\nonumber \\ &=& \frac{3}{4\pi} 
\int dt \,\left|\frac{\partial}{\partial t}\psi^{(1)}_{22}(t,r) \right|^2
\nonumber \\ 
&=& \frac{3}{4\pi}  |A|^2 |\omega_{22}^{(1)}|^2 
\int dt \,e^{2 \Im(\omega_{22}^{(1)}) (t-r_*)} 
\nonumber \\ 
&=& \frac{3}{8\pi}  |A|^2 |\omega_{22}^{(1)}|^2 
\frac{1}{|\Im(\omega_{22}^{(1)})|} \,.
\label{eq:1st-E}
\end{eqnarray}

For the second perturbative order, we also have 
the following formula for the power of gravitational waves 
from the result of the second order gravitational waveform 
in Eq.~(\ref{eq:2ndwave}), 
\begin{eqnarray}
P^{(2)} &=& \frac{45}{8\pi}\,r^2 \sum_{m=\pm4}
\left|\frac{\partial}{\partial t}G_{4m}^{(2)AF}(t,r)\right|^2
\nonumber \\ &=& \frac{45}{4\pi}\, 
\Biggl|\,
\chi^{(2)\,reg}_{4 4}(t,r) + 
{\frac {\sqrt{70}}{1512\sqrt {\pi }}}\,
\Biggl(
27\,\psi^{(1)}_{2 2}(t,r) 
{\frac {\partial^{2}}{\partial t^{2}}}\psi^{(1)}_{2 2}(t,r)
+24\,\left( {\frac {\partial}{\partial t}}\psi^{(1)}_{2 2}(t,r)  \right) ^{2}
\nonumber \\ && \qquad 
+4\,M\,\left({\frac {\partial}{\partial t}}\psi^{(1)}_{2 2}(t,r)\right) 
{\frac {\partial^{2}}{\partial t^{2}}}\psi^{(1)}_{2 2}(t,r)
\Biggr) \,\Biggr|^2
 \,. 
\label{eq:2ndP}
\end{eqnarray}
Here, we have calculated the angular integration 
to derive the first line, and used the relation that 
the second order metric components 
of the $m=4$ and $-4$ modes are complex conjugate, 
i.e., $\chi^{(2)\,reg}_{4 4}=\chi^{(2)\,reg\,*}_{4 -4}$. 
This complex conjugate relation can be shown 
with Eqs.~(\ref{eq:1stSET0}) and (\ref{eq:Zeqchi-reg}). 
We should note that when the first and second order perturbations 
belong to the same harmonics ($\ell,\,m$) mode, 
there are the cross-term contributions 
even after the angular integration, such as 
$G_{\ell m}^{(1)AF} \times G_{\ell m}^{(2)AF\,*}$ 
and $G_{\ell m}^{(1)AF\,*} \times G_{\ell m}^{(2)AF}$ from Eq.~(\ref{eq:dPdO}). 
However, we do not have these terms in our case
that $G_{\ell m}^{(1)AF}$ and $G_{\ell m}^{(2)AF}$
have different harmonics ($\ell,\,m$).

\section{Numerical Calculation: Leaver's method for second-order QNMs}\label{sec:num}

In this section, we consider a numerical method to obtain 
the particular solution $\chi_{44}^{(2)\,reg}(t,r)$ 
for the regularized second order Zerilli equation 
in Eq.~(\ref{eq:Zeqchi-reg}). 
Here we only consider 
the $\ell=4$, $m=4$ mode for the second order calculation 
because $\chi^{(2)\,reg}_{4 -4}$ may be obtained by 
$\chi^{(2)\,reg}_{4 -4}=\chi^{(2)\,reg\,*}_{4 4}$. 
In order to solve the second-order Zerilli equation 
with several digits, we need to know many digits 
of the first order QNM frequencies. 
These are given in Table~\ref{tab:QNM}. 

\begin{table}
\caption{The first order QNM frequencies for the $\ell=2$ mode. 
We have calculated them with higher-precision 
because it is necessary to calculate the second perturbative order.
The QNM frequencies do not depend on $m$ 
for a Schwarzschild black hole.}
\label{tab:QNM}
\begin{center}
\begin{tabular}{lcc}
\hline
\hline
$n$ & $\Re \omega_{2 m}^{(1)}$ & $\Im \omega_{2 m}^{(1)}$ \\
\hline
0 & 0.37367168441804183579349200298 & -0.08896231568893569828046092718 \\
1 & 0.34671099687916343971767535973 & -0.27391487529123481734956022214 \\
2 & 0.30105345461236639380200360888 & -0.47827698322307180998418283072 \\
\hline
\end{tabular}
\end{center}
\end{table}

The following method is basically a modified version 
of the Leaver's continued fraction method~\cite{leaver85}. 
We will finally calculate the second order amplitude 
at infinity and the horizon as the following forms, 
\beqa
\chi_{44}^{(2)\,reg}(t,r=\infty)
&=&C_{I} \left[\omega \psi_{22}^{(1)}(t,r=\infty)\right]^2 \,,
\label{eq:CI}
\\
\chi_{44}^{(2)\,reg}(t,r=2M)
&=&C_{H} \left[\omega \psi_{22}^{(1)}(t,r=2M)\right]^2 \,,
\label{eq:CH}
\eeqa
where $\omega$ denotes the first order QNM frequency $\omega_{2 2}^{(1)}$
and we consider not only the fundamental $n=0$ mode, but also 
the overtone $n=1$ or $2$. 
And then, we may calculate 
the energy of the second order QNM as a function 
of the first order one, 
\beqa
\frac{E^{(2)}}{M}=C_{E} M |\Im \omega| \left(\frac{E^{(1)}}{M}\right)^2 \,,
\label{eq:CE}
\eeqa
where Eq.~(\ref{eq:2ndP}) gives 
\beqa
C_E &=& 20\pi \left|C_I
-\frac{\sqrt{70}}{1512\sqrt{\pi}}
\left(51 - 4 i M \omega \right)
\right|^2 \,.
\eeqa
The coefficients, $C_I$, $C_H$ and $C_E$, are summarized in Table~\ref{tab:C}. 

By Fourier transforming, 
\beqa
\psi_{22}^{(1)}(t,r)&=&\int d\omega \psi_{22\omega}^{(1)}(r) e^{-i\omega t} \,,
\cr
\chi_{44}^{(2)\,reg}(t,r)&=&
\int d\omega \chi_{44\omega}^{(2)\,reg}(r) e^{-i \omega t} \,,
\eeqa
we may solve the second order Zerilli equation with a source term, 
\beqa
\left[\frac{d^2}{d r_{*}^2}+\left(\omega^{(2)}\right)^2 -V_Z(r)\right]
\chi_{44 \omega^{(2)}}^{(2)\,reg}(r)
=S_{44\omega^{(2)}}^{reg}(r) \,,
\label{eq:Z2nd}
\eeqa
where $\omega^{(2)}=2\omega$ 
($\omega^{(2)}_{44}=2\omega^{(1)}_{22}$; $n=0$, $1$ or $2$) 
and the source term is given by 
\begin{eqnarray}
S^{reg}_{44\omega^{(2)}}(r)&=&{\frac {i\omega}{126}}\,\sqrt {\frac{70}{\pi}} 
\Biggl\{  
\Biggl[ 
-{\frac {r \left(7\,r + 4\,M \right) {\omega}^{4}}{r-2\,M}}
\nonumber\\
&& 
+3\,{\frac { \left( 276\,{r}^{7} +476\,{r}^{6}M-1470\,{r}^{5}{M}^{2}-1389\,{r}^{4}{M}^{3}
-816\,{r}^{3}{M}^{4}-800\,{r}^{2}{M}^{5} -555\,r{M}^{6}-96\,{M}^{7}
\right) {\omega}^{2}}{{r}^{3} \left( 3\,r-M \right) ^{2} 
\left( 2\,r+3\,M \right) ^{2} \left( r-2\,M \right) }}
\nonumber\\
&&+ \frac{9\left( r-2\,M \right)}
{{r}^{7} \left( 3\,r+M \right) ^{2} \left( 2\,r+3\,M \right) ^{4}}
\,\left( 2160\,{r}^{9}+11760\,{r}^{8}M
+30560\,{r}^{7}{M}^{2}+41124\,{r}^{6}{M}^{3}+31596
\,{r}^{5}{M}^{4}
\right.
\nonumber\\
&& \left. + 11630\,{r}^{4}{M}^{5}-1296\,{r}^{3}{M}^{6}
-4182\,{r}^{2}{M}^{7}-1341\,r{M}^{8}-144\,{M}^{9} \right)
\Biggr]
\left( \psi_{22\omega}^{(1)} ( r)  \right) ^{2}
\nonumber\\
&&+ \Biggl[ 
-4\,{\frac { \left( r-2\,M \right) ^{2}{M}^{2}{\omega}^{2}}{{
r}^{2} \left( 3\,r+M \right) ^{2}}}
+ \frac{6 \left( r-2\,M \right)}
{{r}^{6} \left( 3\,r+M \right) ^{2} \left( 2\,r+3\,M \right) ^{3}}
\,\left( 144\,{r}^{8}+4116\,{r}^{7}M+2154\,{r}^{6}{M}^{2}
\right.
\nonumber\\
&& \left. - 2759\,{r}^{5}{M}^{3}-8230\,{r}^{4}
{M}^{4}-9512\,{r}^{3}{M}^{5}-3540\,{r}^{2}{M}^{6}-1119\,r{M}^{7}
-144\,{M}^{8} \right)
\Biggr]
\left( {\frac {d}{dr}}\psi_{22\omega}^{(1)}(r)  \right) 
\psi_{22\omega}^{(1)}(r) 
\nonumber\\
&&+ \Biggl[ -{\frac { \left( r-2\,M \right)  
\left( 7\,r+4\,M \right) {\omega}^{2}}{r}}
- \frac{3 \left( r-2\,M \right)}
{{r}^{5} \left( 3\,r+M \right) ^{2} \left( 2\,r+3\,M \right) ^{2}}
\,\left( 228\,{r}^{7}+8\,{r}^{6}M-370\,{r}^{5}{M}^{2}
\right.
\nonumber\\
&& \left. 
+142\,{r}^{4}{M}^{3}-384\,{r}^{3}{M}^{4}
-514\,{r}^{2}{M}^{5}-273\,r{M}^{6}-48\,{M}^{7} \right)  
\Biggr]
\left( {\frac {d}{dr}}\psi_{22\omega}^{(1)}(r)  \right) ^{2} \Biggr\} \,,
\label{eq:2ndZS}
\end{eqnarray}
where $\omega=\omega^{(1)}_{22}$. 

We solve the above equation under the boundary conditions 
with purely ingoing at the horizon and purely outgoing at infinity. 
This gives the two-point boundary value problem. 
However a simple numerical integration like a Runge-Kutta method 
does not work because the QNM frequency has a negative imaginary part 
and hence the QNMs diverge exponentially towards infinity and the horizon. 
We need unrealistic precision to resolve the unwanted solution, 
which is decreasing exponentially as we approach the boundaries. 
Furthermore, since the infinity is an irregular singularity~\cite{leaver85},
the expansion of the basic equation around boundaries 
does not behave well, 
although this method was applied to the first-order~\cite{cd75}.
In the first-order case the most famous method to solve 
this problem is the Leaver's continued fraction method~\cite{leaver85}. 
In the following we modify the Leaver's method and apply it 
to the second-order.

In order to use the Leaver's method, we need to transform the Zerilli 
equation to the Regge-Wheeler one, and vice versa. 
This is possible by the Chandrasekhar transformations~\cite{chandra75} 
for general ($\ell,\,m$) modes, 
\beqa
\psi_{\ell m \omega}^{(1)}(r)
&=&\frac{1}{\frac{4}{9}\lambda^2(\lambda+1)^2+4M^2 \omega^2}
\left(\left(\frac{2}{3}\lambda(\lambda+1)+
\frac{6M^2(r-2M)}{r^2(\lambda r+3M)}\right)
\psi_{\ell m \omega}^{(1) RW}(r)
+2M\frac{d}{dr_*}\psi_{\ell m \omega}^{(1) RW}(r)\right) \,,
\label{eq:chandra1}
\\
\psi_{\ell m \omega}^{(1) RW}(r)&=&
\left(\frac{2}{3}\lambda(\lambda+1)+\frac{6M^2(r-2M)}{r^2(\lambda r+3M)}\right)
\psi_{\ell m \omega}^{(1)}(r)-2M\frac{d}{dr_*}\psi_{\ell m \omega}^{(1)}(r) \,.
\label{eq:chandra2}
\eeqa
We also have similar equations for $\chi_{44 \omega^{(2)}}^{(2)\, reg}(r)$ 
with replacing $\omega$ with $\omega^{(2)}=2\omega$ 
($\omega^{(2)}_{44}=2\omega^{(1)}_{22}$; $n=0$, $1$ or $2$). 
With the Chandrasekhar transformations, 
the Zerilli equation~(\ref{eq:Z2nd}) becomes the Regge-Wheeler one, 
\beqa
\left[\frac{d^2}{d r_{*}^2}+\left(\omega^{(2)}\right)^2 -V_{RW}(r)\right]
\chi_{44 \omega^{(2)}}^{(2) \,reg, RW}(r)
&=&S_{44\omega^{(2)}}^{reg, RW}(r) \,;
\nonumber \\
V_{RW}(r)&=&2 \left(1-\frac{2M}{r}\right)
\frac{(\lambda+1) r-3 M}{r^3} \,,
\label{eq:RWeqS}
\eeqa
where the source term $S_{44\omega^{(2)}}^{reg, RW}(r)$
is also transformed 
by the Chandrasekhar transformations, 
but has a similar form as $S_{44\omega^{(2)}}^{reg}(r)$ 
in Eq.~(\ref{eq:2ndZS}), 
\begin{eqnarray*}
S_{44\omega^{(2)}}^{reg, RW}(r)=
(\cdots) \left(\psi_{22 \omega}^{(1)}(r)\right)^2
+(\cdots)\, \psi_{22 \omega}^{(1)}(r) \left(\frac{d\psi_{22 \omega}^{(1)}(r)}{dr}\right)
+(\cdots) \left(\frac{d\psi_{22 \omega}^{(1)}(r)}{dr}\right)^2 \,.
\end{eqnarray*}
Here we can obtain the first order wave-function $\psi_{22 \omega}^{(1)}(r)$ 
from $\psi_{22 \omega}^{(1) RW}(r)$ with Eq.~(\ref{eq:chandra1}). 

The Leaver's method provides the first order 
Regge-Wheeler function $\psi_{\ell m \omega}^{(1) RW}(r)$ in the form, 
\beqa
\psi_{22 \omega}^{(1) RW}(r)
= 2 M A_{\psi}(r)
\sum_{n=0}^{\infty} a_n \left(\frac{r-2M}{r}\right)^{n} \,,
\eeqa
where
\beqa
A_{\psi}(r)&=&\left(\frac{r}{2M}-1\right)^{\rho} 
\left(\frac{r}{2M}\right)^{-2\rho} e^{-\rho (r-2M)/(2M)} \,;
\nonumber \\
\rho&=&-2 i M \omega \,.
\label{eq:psi-form}
\eeqa
The coefficients $a_n$ are determined by three-term recurrence
relations, 
\beqa
\alpha_0 a_1+\beta_0 a_0&=&0 \,,
\nonumber \\
\alpha_n a_{n+1}+\beta_n a_n+\gamma_n a_{n-1}&=&0 \,,
\quad n=1,\,2,\,\dots ,
\eeqa
where $a_0$ is arbitrary and $\alpha_n$, $\beta_n$ and $\gamma_n$ 
are given by 
\beqa
\alpha_n&=&n^2+(2\rho+2)n+2\rho+1 \,,
\nonumber\\
\beta_n&=&-(2n^2+(8\rho+2)n+8\rho^2+4\rho+2\lambda-1) \,,
\nonumber\\
\gamma_n&=&n^2+4\rho n+4\rho^2-4 \,.
\label{eq:abc}
\eeqa

Similarly the second order solution can be written in the form, 
\beqa
\chi_{44\omega^{(2)}}^{reg,RW}(r)
=\left[A_\psi(r)\right]^2
\sum_{n=0}^{\infty} a_{n}^{(2)} \left(\frac{r-2M}{r}\right)^n \,.
\eeqa
The coefficients $a_n^{(2)}$ are determined by the following 
three-term recurrence relations, 
\beqa
\alpha_0 a_1^{(2)}+\beta_0 a_0^{(2)}&=&b_0 \,,
\nonumber \\
\alpha_n a_{n+1}^{(2)}+\beta_n a_n^{(2)}+\gamma_n a_{n-1}^{(2)}
&=&b_n \,, \quad n=1,\,2,\,\dots \,,
\eeqa
where we replace $\rho$ with $2\rho$ in $\alpha_n$, $\beta_n$ and
$\gamma_n$ in Eqs.~(\ref{eq:abc}) and the source terms $b_{n}$ are 
determined by expanding $S^{reg,RW}_{44\omega^{(2)}}$ in Eq.~(\ref{eq:RWeqS}) as
\beqa
\frac{r^3}{r-2M}S^{reg,RW}_{44\omega^{(2)}}(r)
=\left[A_\psi(r)\right]^2
\sum_{n=0}^{\infty} b_{n} \left(\frac{r-2M}{r}\right)^n \,.
\eeqa
We can first obtain the source terms $b_n$ algebraically, 
and then evaluate them numerically with the {\it Maple} calculator.
Once we have numerical values of $b_n$, we can solve the recurrence
relations. 
The solution that satisfies the boundary condition at infinity
is determined by adjusting $a_0^{(2)}$ 
for which $\sum a_n^{(2)}$ exists and is finite. 

\begin{table}
\caption{The coefficients for the amplitude and energy of 
the second order QNMs 
in Eqs.~(\ref{eq:CI}), (\ref{eq:CH}) and (\ref{eq:CE}). 
The first order QNM has $\ell=2, m=2$ with the overtone $n$.}
\label{tab:C}
\begin{center}
\begin{tabular}{lccc}
\hline
\hline
$n$ & $C_I$ & $C_H$ & $C_E$ \\
\hline
0 & $0.221-0.489i$ & $-0.221+1.19i$ & $15.0$ \\
1 & $0.397-0.384i$ & $0.262-1.16i$ & $12.7$ \\
2 & $0.459-0.226i$ & $0.604-0.811i$ & $8.97$ \\
\hline
\end{tabular}
\end{center}
\end{table}

\section{Summary and Discussions}\label{sec:dis}

In this paper, we have investigated 
the second order QNMs of a Schwarzschild black hole, 
as summarized in the following.
\begin{itemize}
\item[1.] We consider the $\ell=2, m=\pm 2$ even parity modes 
with the QNM frequencies $\omega_{2 \pm2}^{(1)}$ 
for the first order perturbations
because these modes dominate for binary black hole mergers. 
\item[2.] We have considered the $\ell=4, m=\pm 4$ even parity modes 
for the second perturbative order that are driven by 
the first order perturbations.
We have employed the second order Regge-Wheeler-Zerilli formalism 
with the source term written 
by quadratic terms of the first order wave-function. 
\item[3.] We have regularized the source term 
because this does not behave well at the boundaries. 
\item[4.] Based on the regularized source, 
the second order wave-function have been obtained 
by using the modified Leaver's continued fraction method. 
\item[5.] We have explicitly derived the gauge transformation
of the metric perturbation into an asymptotic flat gauge
to extract physical information 
from the first and second order wave-functions. 
We have also formulated the power of gravitational waves 
for the first and second perturbative order. 
\item[6.] As the result, the second order QNM frequencies are found 
to be $\omega^{(2)}_{4\pm4}=2\omega^{(1)}_{2\pm2}$ 
and the gravitational wave amplitude could go up to $\sim 10\%$. 
This means that we can in principle detect and identify the higher order QNMs
since their frequencies differ from any first order ones. 
The detectability of a two-mode ringdown wave has been discussed 
in~\cite{Berti:2007zu}. 
\end{itemize}

The detection of the second order QNMs would have the following advantages. 
\begin{itemize}
\item[1.] The second order QNMs would 
provide a new test of general relativity,
in particular of the no-hair theorem.
According to the no-hair theorem,
the astrophysical BHs are completely characterized by
their mass and angular momentum.
Then the mass and angular momentum derived from the second order QNMs
should coincide with that from the first order ones.
\item[2.] The first and second order QNM amplitudes 
include information about the total radiated GW energy 
$E^{(1)}$ and $E^{(2)}$, respectively. 
Using these ratio, this could provide distance indicators 
from a data analysis of only the QNM gravitational waves, 
since the observed GW amplitude is $h \sim (E/M)^{1/2}(M/r)$. 
This is important for the case that
the inspiralling phase is out of the detector frequency range.
To be precise, since the observed GW amplitude also depends 
on the zenith angle $\theta$, the parameters are degenerated
and we can determine only the order-of-magnitude distances.
A possible solution to this problem is presented below.
\item[3.] We may also use the first-to-second amplitude ratio 
to reject fake events in the QNM search in which there are many fake 
events~\cite{Tsunesada:2005fe}. 
In \cite{Tsunesada:2005fe}, 
a method to remove fake events has been proposed 
by using the overtone QNM of the same ($\ell,\,m$) mode. 
The amplitude of the overtone QNM is
determined by two factors, the `excitation coefficient' of the QNM 
and the `initial condition' of a perturbation. 
We can derive the excitation coefficient 
theoretically~\cite{leaver86,Andersson:1996cm,Andersson:1999wj,
Glampedakis:2001fg,Glampedakis:2003dn}, 
but we must give the initial condition which depends on 
how QNMs are excited.
Thus, it has been difficult to remove fake events 
by using the overtone QNM with an undetermined amplitude.
On the other hand,
the second order QNMs have the predictable amplitude
except for the zenith angle dependence.
This may ease the fake event rejection.
\end{itemize}

Future problems include
\begin{itemize}
\item[1.] It is necessary to formulate the odd parity mode case 
in the Schwarzschild background. 
The odd parity mode appears when BHs have spin before mergers. 
\item[2.] We also need to discuss coupling contributions 
between different harmonics ($\ell,\,m$). 
\item[3.] As discussed in Sec.~\ref{sec:2nd},
the product of $\ell=2, m=\pm 2$ even modes
gives not only the $m=\pm 4$ modes, which is studied in this paper,
but also the $m=0$ mode.
Although the $m=0$ mode does not oscillate as a function of time,
the amplitude would be comparable to that of the $m=\pm 4$ modes.
Since the metric perturbation of the $m=0$ mode
does not return to the initial value,
this mode represents the gravitational memory effect
\cite{Christodoulou:1991cr,thorne92}.
If the $m=0$ mode can be also detected,
the amplitude ratio of the $m=0$ mode to the $m=\pm 4$ modes
would determine the zenith angle to the observer
since the zenith angle dependence is different.
This could resolve the parameter degeneracy in
the distance measurements with the second order QNMs
discussed above.
\item[4.] We have to extend the analysis to the Kerr BH case. 
When BHs have no spin before mergers, 
the final spin of the remnant BH is $a \sim 0.7$~\cite{pre05}
and hence the Kerr effects may not be so large 
as inferred from the fact that the QNM frequencies shift 
by only a small factor. 
However, the final spin becomes large 
in the case of highly-spinning BH binaries~\cite{Campanelli:2006fy}. 
Since the master equation for Kerr BHs also has a source term 
that is quadratic in the first order function~\cite{Campanelli:1998jv}, 
we may expect similar results. 
\item[5.] It is also important to discuss a mathematically rigorous 
definition of second order QNMs like 
the first order ones that use the Laplace transformation
rather than the Fourier transformation~\cite{leaver86,kokkotas99}. 
\item[6.] The third order formulation is interesting. 
As suggested by \cite{landau76}, 
the QNM frequencies will also blueshift up to 
$(\psi^{(1)}/M)^2\sim 1\%$ at this third order. 
\item[7.] In order to proof that the second-order QNMs actually exist
and stand out of the GW tail,
we have to find the second-order QNMs directly
in the numerical simulations.
Such simulations are challenging because the mesh size
should be less than $\sim 1\% \times M$
to resolve $\sim 1\%$ metric perturbations.
There are detail analyses of nonlinear mode-coupling effects 
on a waveform by using the full nonlinear 
code~\cite{Baker:1999sj,Zlochower:2003yh}, 
but the second-order QNMs have not been identified. 
\end{itemize}

\acknowledgments

We would like to thank Carlos~O.~Lousto for his helpful advice and 
warm encouragement. 
H.N. is supported by JSPS Postdoctoral Fellowships 
for Research Abroad and in part by 
the NSF for financial support from grant PHY-0722315.
This work is also supported in part by
the Grant-in-Aid (18740147) from the 
Ministry of Education, Culture, Sports, Science and Technology
(MEXT) of Japan (K.I.).


\appendix 

\section{QNM frequencies}\label{app:qnm}

In this appendix, we consider quasi-normal frequencies of 
a black hole in the first perturbative order 
in order to clarify their complex nature and 
relation between positive and negative $m$ modes. 
First, we discuss the quasi-normal frequencies of 
a Kerr black hole, and then go back to the Schwarzschild case. 

With the spacetime symmetry, 
we may use the spheroidal (spherical in the Schwarzschild case) 
harmonics expansion which has the labels, $\ell$ and $m$. 
There are infinite overtone frequencies of the QNM 
for each $(\ell,\, m)$ mode. 
This is labeled by $n$. 
Even if we fix $\ell$, $m$ and $n$,
there exist two quasi-normal frequencies 
as we can see from Fig.~1 and 3 in~\cite{leaver85}. 
One has a positive real part $\omega_{\ell m n (+)}$ 
while the other has a negative one $\omega_{\ell m n (-)}$. 

For $m \ge 0$, the QNM frequency 
with a positive real part, $\omega_{\ell m n (+)}$, represents
more slowly damped modes than that with a negative real part,
$\omega_{\ell m n (-)}$, while for $m \le 0$, 
$\omega_{\ell m n (-)}$ represents 
more slowly damped modes than $\omega_{\ell m n (+)}$.
Since we concentrate on the most dominant modes, 
we consider $\omega_{\ell m n (+)}$ 
and $\omega_{\ell -m n (-)}$ ($m \geq 0$) in this paper. 
For example, although the integration in Eq.~(\ref{eq:fourier-psi})
picks up the two set of frequencies, $\omega_{\ell m n (+)}$ 
and $\omega_{\ell m n (-)}$, 
we adopt the most slowly damped mode as 
\begin{eqnarray}
\psi_{\ell m}(t,r) &=& e^{-i\omega_{\ell m 0 (+)} t} 
\psi_{\ell m\omega_{\ell m 0 (+)}}(r) 
+ e^{-i\omega_{\ell m 0 (-)} t} \psi_{\ell m\omega_{\ell m 0 (-)}}(r) 
\label{eq:formP}
\\
&\sim &\left\{
\begin{array}{ll}
e^{-i\omega_{\ell m 0 (+)} t} \psi_{\ell m\omega_{\ell m 0 (+)}}(r) \,,
& \quad {\rm for} \quad m\ge 0 \,,\\
e^{-i\omega_{\ell m 0 (-)} t} \psi_{\ell m\omega_{\ell m 0 (-)}}(r) \,,
& \quad {\rm for} \quad m<0 \,,
\end{array}\right.
\end{eqnarray}
where we consider the fundamental $n=0$ mode. 
We note that the QNM frequencies have the complex conjugate symmetry, 
\begin{eqnarray}
\omega_{\ell m n (+)} &=& -\omega_{\ell -m n (-)}^* \,,
\label{eq:QNMsym}
\end{eqnarray}
as discussed in Sec.~4 of \cite{leaver85}. 

Next, we discuss the Schwarzschild case. 
In this case, the $m$ modes degenerate 
due to the spherical symmetry. Therefore, 
the label $m$ have no meaning and we may write 
\begin{eqnarray}
\omega_{\ell m n (+)} &=& -\omega_{\ell m n (-)}^* \,.
\nonumber 
\end{eqnarray}
This equations means that the frequencies $\omega_{\ell m n (+)}$ 
and $\omega_{\ell m n (-)}$ have the same damping time. 
The wave-function $\psi_{\ell m}(t,r)$ of the fundamental QNM is 
formally written by the same equation in Eq.~(\ref{eq:formP}). 
The two terms in the right hand side of Eq.~(\ref{eq:formP}) 
have the same damping time in the Schwarzschild case, 
but we do not consider 
the term with $\omega_{\ell m n (-)}$ for a positive $m$ mode 
and $\omega_{\ell m n (+)}$ for a negative $m$ mode, 
because these modes are not the more slowly damped mode 
in the Kerr case. Therefore, 
we have adopted the following wave-function 
for the first perturbative order, 
\begin{eqnarray}
\psi_{\ell m}(t,r) &=& e^{-i\omega_{\ell m}^{(1)} t} 
\psi_{\ell m \omega_{\ell m}^{(1)}}(r) \,, \quad {\rm for} \ \ m \geq 0 \,,
\nonumber \\ 
\psi_{\ell m}(t,r) &=& e^{i \omega_{\ell m}^{(1)*} t} 
\psi_{\ell m -\omega_{\ell m}^{(1)*}}(r) \,, \quad {\rm for} \ \ m < 0 \,,
\label{eq:1stSET}
\end{eqnarray}
where $\omega_{\ell m}^{(1)}=\omega_{\ell m n (+)}$ 
and we consider $n=0$, $1$ or $2$ for comparison in this paper. 
These wave-functions satisfy Eq.~(\ref{eq:1stSET0})
ensuring real metric components.


\end{document}